\documentclass[final,onefignum,onetabnum]{siamonline250211}

\usepackage{lipsum}
\usepackage{graphicx}
\usepackage{epstopdf}
\usepackage{algorithm}
\usepackage{amsmath,amssymb,amsfonts,mathrsfs}
\usepackage{stmaryrd}
\usepackage{textcomp}
\usepackage{stfloats}
\usepackage{url}
\usepackage{verbatim}

\usepackage{microtype}
\usepackage{booktabs}
\usepackage{hyperref}
\usepackage{mathtools}
\usepackage[noend]{algpseudocode}
\usepackage{multirow}
\usepackage{bm}
\usepackage{color}
\usepackage{wrapfig}
\usepackage[textsize=tiny]{todonotes}
\usepackage{multicol}
\usepackage{subfig}

\def\lbrk{\bm \llparenthesis}
\def\rbrk{\bm \rrparenthesis}
\def\setsep{\,|\,}

\def\sph2{\mathbb{S}^{2}}
\def\sp1{\mathbb{S}^{1}}

\def\Span{\mathrm{span}}

\DeclareMathOperator{\Tr}{Tr}

\makeatletter
\def\namedlabel#1#2{\begingroup
    #2%
    \def\@currentlabel{#2}%
    \phantomsection\label{#1}\endgroup
}
\makeatother

\ifpdf
  \DeclareGraphicsExtensions{.eps,.pdf,.png,.jpg}
\else
  \DeclareGraphicsExtensions{.eps}
\fi

\usepackage{enumitem}
\setlist[enumerate]{leftmargin=.5in}
\setlist[itemize]{leftmargin=.5in}

\newsiamremark{remark}{Remark}
\newsiamremark{hypothesis}{Hypothesis}
\crefname{hypothesis}{Hypothesis}{Hypotheses}
\newsiamthm{claim}{Claim}
\newsiamremark{fact}{Fact}
\crefname{fact}{Fact}{Facts}

\headers{Data-Adaptive Graph Framelets}{R. Zheng and X. Zhuang}

\title{Data-Adaptive Graph Framelets with Generalized Vanishing Moments for Graph Machine Learning\thanks{Submitted to the editors DATE.
\funding{This work was supported in part  by the Research Grants Council of Hong Kong (Project no. CityU 11309122, CityU 11302023, CityU 11301224, and CityU 11300825).}}}

\author{Ruigang Zheng\thanks{Department of Mathematics, City University of Hong Kong, Tat Chee Avenue, Kowloon Tong, Hong Kong (\email{ruigzheng2-c@my.cityu.edu.hk}.}
\and Xiaosheng Zhuang\thanks{Department of Mathematics, City University of Hong Kong, Tat Chee Avenue, Kowloon Tong, Hong Kong  (\email{xzhuang7@cityu.edu.hk}.}
}

\usepackage{amsopn}

\ifpdf
\hypersetup{
  pdftitle={Data-Adaptive Graph Framelets},
  pdfauthor={R. Zheng and X. Zhuang}
}
\fi

\begin{document}

\maketitle

\begin{abstract}
In this paper, we propose a general framework for constructing tight framelet systems on graphs with localized supports based on partition trees. Our construction of framelets provides a simple and efficient way to obtain the orthogonality with $k$ arbitrary orthonormal vectors. When the $k$ vectors contain most of the energy of a family of graph signals, the orthogonality of the framelets intuitively possesses ``generalized ($k$-)vanishing'' moments, and thus, the coefficients are sparse. Moreover, our construction provides not only framelets that are overall sparse vectors but also fast and schematically concise transforms. In a data-adaptive setting, the graph framelet systems can be learned by conducting optimizations on Stiefel manifolds to provide the utmost sparsity for a given family of graph signals. Furthermore, we further exploit the generality of our proposed graph framelet systems for heterophilous graph learning, where graphs are characterized by connecting nodes mainly from different classes. The usual assumption that connected nodes are similar and belong to the same class for homophilious graphs is contradictory for heterophilous graphs. Thus, we are motivated to bypass simple assumptions on heterophilous graphs and focus on generating rich node features induced by the graph structure, so as to improve the graph learning ability of certain neural networks in node classification. We derive a specific system of graph framelets and propose a heuristic method to select framelets as features for neural network input. Several experiments demonstrate the effectiveness and superiority of our approach for non-linear approximation, denoising, and node classification.
\end{abstract}



\section{Introduction}\label{sec:intro}

\subsection{Background and Motivation}\label{intro_subsec_1}
Graphs are prevalent data formats that have a variety of realizations in real life, e.g., social networks, traffic networks, sensor networks, etc, and the area of graph machine learning including graph signal processing (GSP), node classification, link prediction, and so on, has been one of the most active research areas in the past decade \cite{shuman2013emerging,ortega2018graph, wu2020comprehensive}. Given its practical importance, it is desired to have efficient processing tools on graphs. Compared with signals, images, and other Euclidean-type data, graph data/signals are more irregular and flexible. Therefore, classical signal processing methods cannot be applied directly to such data. In graph signal processing, a central topic is to develop notions and tools that resemble those on the continuous spaces or the discrete $d$-dimensional ($d$-D) signals, e.g., the Fourier and wavelet/framelet transforms. Stemmed from the spectral graph theory \cite{chung1997spectral}, the graph Fourier transforms (GFTs) based on graph Laplacians are by far one of the most fundamental concepts/tools upon which many other tools are developed.

Sparsity is one of the key characteristics in achieving efficient and effective signal processing since it implies the capability of representing signals with significantly fewer coefficients. A typical way to achieve sparsity in classical $d$-D discrete wavelet/framelet transforms is to construct high-pass filters with high order of vanishing moments \cite[Sec. 1.2]{han2017framelets}. To elaborate, since smooth functions can be approximated by polynomials (using the Taylor expansion), polynomial sequences convolving with a filter with high order of vanishing moments vanish up to a certain order, and thus, the wavelet/framelet (high-pass filter) coefficients of the discrete samples of smooth functions are consequently sparse. The theory for classical wavelet/framelet analysis is well-established; see, e.g., \cite{chui1992introduction,daubechies1992ten,han2017framelets,mallat2008wavelet}. For more complicated families of signals, wavelets/framelets with more desirable properties than those of the ordinary wavelets, for example, the directionality, are required in order to achieve (optimal) sparsity for the representations of specific families of signals \cite{candes2004new,kutyniok2012introduction,han2015smooth,han2016directional,zhuang2016digital, che2018digital,han2019directional-AML,han2019directional-SIIMS}.

In contrast, it is generally difficult to define the corresponding notions of vanishing moments, smooth functions, and approximation properties for signals on graphs due to the irregular structure of graphs. It is also difficult to define families of signals of graphs with special properties, e.g., directional signals on graphs. Therefore, it is plausible that sparsity is seldom theoretically discussed in graph signal processing. However, there are still some attempts to define generalized vanishing moments for graphs.
In \cite{sharon2015class}, vectors with the property of being orthogonal to the first $k$ graph Laplacian eigenvectors are regarded as having \emph{generalized $k$ vanishing moments}. A graph wavelet system is subsequently defined, and the wavelets are guaranteed to have this property. For signals that are well approximated by the first $k$ eigenvectors, the system is able to produce very sparse coefficients. This inspires us to consider a more general construction of graph framelet systems in which a family of signals is not assumed to be well-approximated by the first $k$ eigenvectors, but still largely lies in a $k$-dimensional subspace. Similarly, the orthogonality with this subspace can also be considered to have generalized vanishing moments, and a system with this property can produce sparse coefficients for the family of signals. More precisely, we focus on the following question:
\begin{itemize}
\item[\namedlabel{itm:q1}{\rm Q1})] Given a family $\mathfrak{F}$ of graph signals that is approximately included in a low-dimensional linear subspace in $\mathbb{R}^n$, how to define orthonormal bases and tight frames that  1) result in sparse matrices if collecting all the (frame) vectors, 2) sparsely represent vectors in $\mathfrak{F}$ by exploiting the low dimension of the linear subspace, and 3) associated with fast forward and backward transforms?
\end{itemize}
To the best of our knowledge, none of the works in the literature have considered our notion of generalized $k$-vanishing moments, and they have at least one of the following issues (see Section \ref{intro_subsec_4}): (a) Do not allow perfect reconstruction; (b) only provide bases and thus are restricted to critical sampling; (c) result in dense matrices; and (d) without fast transforms. 
In this paper, we define general graph framelets that possess the property of having generalized $k$ vanishing moments and do not carry any of the above issues. 


With such general graph framelet systems, we further consider other tasks in graph machine learning that could demonstrate different capabilities of our systems. We specifically consider the task of node classification on heterophilous graphs and focus on the following question: 
\begin{itemize}
\item[\namedlabel{itm:q2}{\rm Q2})] How do we apply the general graph framelet system so that the framelets represent structural information on graphs other than the adjacency matrices?
\end{itemize}
In fact, this is motivated by the special properties of heterophilous graphs. If we inspect the classes of two nodes on a randomly chosen edge, the classes are more likely to be different. This also means that for any node, the majority of its neighbors come from other classes. This is contradictory to the usual assumption that graphs connect similar entities, upon which classical methods such as \cite{zhou2003learning} utilize graph Laplacians to measure smoothness and impose smoothing on the outcomes. Thus, we are motivated to bypass simple assumptions on heterophilous graphs and focus on generating rich node features induced by the graphs. Moreover, these features should be different from the columns of the adjacency matrices, which can be considered as the most primitive structural information. One can expect that, combined with the generated features, neural networks are capable of extracting complex patterns and improving node classification on heterophilous graphs. Such an approach can be considered as feature engineering on graphs, which also appears in some of the latest GNNs (see Section \ref{intro_subsec_5}). 

Before proceeding further, we discuss some related works on graph signal processing and heterophilous graph learning below.

\subsection{Related Works}

\subsubsection{Graph Signal Processing}\label{intro_subsec_4}
By GFTs, notions of frequencies, low-pass filters, and high-pass filters can be defined similarly to those of the Euclidean counterparts. These concepts further facilitate the development of an analysis-synthesis framework using two-channel filter banks and up/down samplings on graphs analogous to classical 1-D cases. The work in \cite{narang2012perfect} shows that this can be established on bipartite graphs, on which up/down sampling is the simple restriction on either set of the bipartition. This is essentially due to the algebraic property of \textit{spectral folding} that is satisfied on the bipartite graphs. For arbitrary graphs, this property is not guaranteed, and one way to tackle this issue is to decompose the original graph into bipartite subgraphs. The graph signal processing is then conducted independently on each bipartite subgraph. Subsequent work improves \cite{narang2012perfect} from the following aspects: (1) Using compact (polynomial) and biorthogonal filters by allowing different filter banks in the analysis and synthesis phases \cite{narang2013compact}. (2) Applying over-sampled bipartite graphs since the algorithm of decomposition in bipartite subgraphs in \cite{narang2012perfect} results in a loss of edges of the original graphs \cite{sakiyama2014oversampled}. (3) Improving computational complexity and quality of down-samplings by using maximum spanning trees \cite{nguyen2014downsampling,zheng2019framework}. (4) Generalizations to arbitrary graphs by applying generalized sampling operators, generalized graph Laplacians, and generalized inner-products \cite{you2023perfect,pavez2023two}. All of these have shown certain capabilities among various graph signal processing tasks.

Apart from the aforementioned work, there are many other realizations of graph signal analysis-synthesis tools based on various motivations and possessing other types of characteristics and merits \cite{li2019scalable,tremblay2016subgraph,sakiyama2019two,kotzagiannidis2019splines,miraki2021spline,miraki2021modified,ekambaram2015spline,you2023spline}. From a general perspective, we can unify these GSP approaches by viewing graph signals as vectors in $\mathbb{R}^n$ and one is aiming at the construction of representation systems for graph signals have various desired properties. The analysis and synthesis phases with respect to a representation system are typically corresponding to two matrices $\bm{T}_a$ and $\bm{T}_s$, respectively. One of the most important properties of a \emph{graph signal representation system} is the \emph{perfect reconstruction} property, i.e., $\bm{T}_s\bm{T}_a = \bm{I}$ with $\bm I$ being the identity matrix of a certain size. The property of critical sampling, i.e., $\bm{T}_a,\bm{T}_s\in \mathbb{R}^{n\times n}$, is also desirable in some applications such as the graph signal compression. The perfect reconstruction allows for recovering the signal after decomposition, while the critical sampling indicates that, compared with the original signal, there is no extra storage overhead after analysis. 

However, critical sampling could be a drawback in some other applications. For example, in graph signal denoising, redundancy plays a crucial role in exploiting the intrinsic structures of the underlying noisy signal. By constructing systems with $\bm{T}_a$ as a redundant frame operator to represent graph signals rather than a non-redundant basis, one can generate smaller coefficients from such a system for noisy signals. Therefore, the portions of noise have a larger chance of being thresholded. Moreover, in terms of tightness of the representation systems, except for special cases in \cite{narang2012perfect,tremblay2016subgraph,sakiyama2019two}, in the other aforementioned work, $\bm{T}_s \neq \bm{T}_a^\top$ (the transpose of $\bm T_a$) and thus we can not represent the analysis-synthesis framework using only $\bm{T}_a$. The tightness of the representation system requires the construction of  $\bm{T}_a$ and $\bm{T}_s$ such that $\bm{T}_s = \bm{T}_a^\top$, $\bm{T}_a^\top\bm{T}_a = \bm{I}$ with $\bm{T}_a \in \mathbb{R}^{m\times n}$ for some $m\geq n$,  which is equivalent to require that the rows of $\bm{T}_a$ forms a tight frame on $\mathbb{R}^n$. There are a few existing works \cite{Hammond2011wavelets,dong2017sparse,behjat2016signal,leonardi2013tight,tay2017almost} on tight frames for graph signal representations, most of which are derived from GFTs and thus are called spectral wavelets/framelets \cite{Hammond2011wavelets}. While spectral wavelets/framelets are well-interpreted in the frequency domain,  the wavelets/framelets generally do not have localized support unless the filters are polynomial-type. In this case, $\bm{T}_a$ is a dense matrix, which leads to computation and storage burdens. To facilitate efficient computations, polynomial approximations to the target filters are applied. However, it fails to be a tight frame, i.e., $\bm{T}^\top_a\bm{T}_a \neq \bm{I}$ \cite{Hammond2011wavelets}. On the other hand, following \cite{crovella2003graph,gavish2010multiscale,chui2015representation,chui2018representation} that utilize a series of hierarchical partitions on graphs, a recent work \cite{li2024permutation} proposes Haar-type graph framelets with localized support.

\subsubsection{Node Classification in Heterophilous Graph Learning}\label{intro_subsec_5}

A typical task on graph learning is to identify the classes to which each node belongs. Such classification corresponds to determining the group an account belongs to in a social network or the type of traffic connection in a traffic network. There are abundant works in the literature that deal with node classification, where graph neural network (GNN) approaches are currently one of the most active topics \cite{wu2020comprehensive}. However, graphs are often assumed to connect similar entities. In node classification, such assumptions mean that connected nodes are more likely to be of the same classes, which is described as being \textit{homophilous}. The datasets adopted in the early works are primarily homophilous, and these GNNs implicitly perform graph smoothing \cite{kipf2016semi,li2018deeper}. In recent years, an ongoing active topic has been the counterpart of homophilous graphs, the so-called \textit{heterophilous} graphs \cite{zheng2022graph}. Intuitively, heterophily means that connected nodes are more likely to be of different classes. As a result, such a reversing nature of heterophilous graphs brings challenges to the GNNs that perform smoothing. Other types of GNNs were therefore proposed to handle node classification on heterophilous graphs.

\par The adaptations for heterophilous graphs in current GNNs are based on two observations: 1) Spatial aspect: For a fixed node, other nodes from the same class are distributed more outside the 1-hop neighborhood. 2) Spectral aspect: Target signals on heterophilous graphs have larger oscillations with respect to the graph Laplacian. Thus, its frequency distribution can hardly be confined to the low-frequency part. GNNs for heterophilous graphs mainly utilize graph-induced aggregations in multiple layers, which resembles the idea in deep convolutional neural networks. These GNNs can be further categorized into two types according to whether the aggregations are spatially \cite{velivckovic2018graph,hamilton2017inductive} or spectrally defined \cite{defferrard2016convolutional,he2021bernnet,zheng2021framelets}. Recently, there has been a third perspective in GNNs, in which the multiple layers of neural networks are simply linear, and all graph-induced aggregations are only involved in forming inputs for the networks \cite{maurya2022simplifying,huang2024how}. These methods aim at jointly generating new features for all nodes. Then, with these new features as inputs, the following training is completely supervised. Such approaches have demonstrated superior performance in heterophilous node classification despite not using complicated neural network architectures. In our view, these methods can be categorized as feature engineering for heterophilous graphs.

\subsection{Contributions}
In the following sections, we propose framelets on graphs that generalize previous work and fit the requirements in \ref{itm:q1}. We will also be looking at a special realization of the graph framelet system. By such an approach, we are able to provide rich, sparse graph framelets as features for node classification. Rich framelets mean that we can choose among various candidate framelets. Being sparse means that when all framelets are regarded as a single matrix, the matrix is sparse. The framelets are constructed upon an induced two-hop graph so as to provide structural information that is different from the original one-hop adjacency matrix. The framelets are then measured by the variance on the original adjacency matrix, intended for integrating one-hop and two-hop structural information. By sorting the variance, we are able to provide a spectral display of the framelets in which certain parts suggest substantial differences among the framelets. These parts are then selected as features for neural network input. The contributions of our proposed graph framelet systems are summarized as follows:
\begin{enumerate}
\item[{i)}] We propose a general method to construct tight frames on graphs, in which the graph framelets have localized support and realizable, fast, and concise transforms.

\item[{ii)}] We propose using our graph framelet systems to adapt to the low-dimensional structure of a given family of graph signals so that the resulting graph framelet systems possess generalized vanishing moments, facilitating sparse representation.

\item[{iii)}] We propose a special realization of the graph framelet system for generating new features as neural network input.

\item[{iv)}] We demonstrate the effectiveness of our proposed method by comparing it with other approaches on non-linear approximation, denoising, and node classification.
\end{enumerate}

\subsection{Outline of the paper}
The rest of this paper is organized as follows. The construction of $\mathcal V$-framelet  and $\mathcal{G}$-framelet systems are described in Section~\ref{sec:frame}. The graph framelet transforms and their computational complexities are discussed in Section~\ref{sec:gmt}.  Data-adaptive realizations of the graph framelet systems based on Stiefel manifold optimizations are detailed in Section \ref{sec:data-adaptive}. In Section \ref{sec:expr}, numerical experiments and comparisons are conducted for the tasks of non-linear approximation and denoising in graph signal processing. Section \ref{sec:gnn} shows the definition of the $2$-hop graph framelet systems and their applications to node classification. Conclusion and final remarks are given in Section \ref{sec:conclude}.

\section{Graph Framelet Systems}\label{sec:frame}
In this section, we focus on the detailed construction of general graph framelet systems and the characterizations of such systems to be tight frames. Moreover, such framelet systems can enjoy desirable properties such as localized supports, generalized vanishing moments, fast transform algorithms, and so on.

\subsection{Preliminaries}
We first introduce some necessary notation, definitions, and preliminary results. 
We denote an undirected weighted graph with $n$ vertices as $\mathcal{G}:=(\mathcal{V},\mathcal{E},\bm{W})$ (or simply $\mathcal G:=(\mathcal V, \mathcal E)$), where $\mathcal{V}:=\{v_1,v_2,\dots,v_n\}$, $\mathcal{E}\subset \mathcal{V}\times \mathcal{V}$, $\bm{W}=(w_{ij})_{1\le i,j\le n}\in \mathbb{R}^{n\times n}$ denote the set of vertices, the set of edges, and the weight (adjacency) matrix, respectively. The space $L^2(\mathcal{G}):=\{f\setsep \mathcal{V} \rightarrow \mathbb{R}\}$ is the collection of graph signals on $\mathcal{G}$ and can be regarded as the finite-dimensional Hilbert space $\mathbb{R}^n$ with the usual Euclidean inner-product $\langle \cdot, \cdot \rangle$ and induced ($l_2$) norm $\Vert \cdot \Vert:=\sqrt{\langle \cdot, \cdot\rangle}$. The cardinality of a set is denoted by $|\cdot|$. Let $\bm{L} := \bm{D} -\bm{W}$ denote the \emph{unnormalized graph Laplacian} matrix, where the diagonal matrix $\bm{D}:= \text{diag}(d_1,d_2,\dots,d_n)$ with  $d_i:= \sum_{j=1}^n w_{ij},1\leq i\leq n$ is the degree matrix. $\bm{L}$ is positive semidefinite and thus it has $n$ eigenvalues $0\leq \lambda_1\leq \lambda_2\leq \dots \leq \lambda_n$ associated with real orthonormal eigenvectors $\bm{u}_1,\bm{u}_2,\dots,\bm{u}_n$. If $\mathcal{G}$ is connected, then $\bm{u}_1$ a vector with constant element $1/\sqrt{n}$. Let $\tilde{\bm{L}}:= \bm{I} - \bm{D}^{-1/2}\bm{W}\bm{D}^{-1/2}$ denote the \emph{normalized graph Laplacian} matrix. The quadratic form $\bm{f}\tilde{\bm{L}}\bm{f}^\top, \bm{f}\in \mathbb{R}^{1\times n}$ equals
\begin{equation}\label{ch4_eq_1}
	\sum_{e =(v_i,v_j)\in \mathcal{E}} \frac{w_{ij}(\bm{f}_i -\bm{f}_j)^2}{\sqrt{\bm{D}_{ii}\bm{D}_{jj}}}, 
\end{equation}
which measures the variance of a signal $\bm{f}$ on a graph by summing (normalized) value differences on all edges. For unit-norm $\bm{f}$, $\bm{f}\bm{L}\bm{f}^\top \in [0,2]$ \cite{huang2024how}.

For $K\in \mathbb{N}$, the set $\{1,\dots,K\}$ is denoted as $[K]$. For a matrix $\bm{M}$, we denote its $i$-th row as $\bm{M}_{i,:}$. We denote $\bm I$ the identity matrix of a certain size and omit its size for simplicity. In this paper, vectors in $\mathbb{R}^d,d\in \mathbb{N}$ are assumed to be {\bf row vectors}, or equivalently matrices in $\mathbb{R}^{1\times d}$. 
Given a set $\{\bm{X}_i\}_{i=1}^m \subseteq \mathbb{R}^{s\times t}$ of $m$ matrices, we use $\mathrm{CONCAT}(\{\bm{X}_i\}_{i=1}^m):=\llbracket\bm X_i\rrbracket_{i\in[m]}:=[\bm X_1^\top,\ldots, \bm X_m^\top]^\top$  to denote the matrix in $\mathbb{R}^{ms\times t}$ constructed by concatenating $\bm{X}_1,\bm{X}_2,\dots,\bm{X}_m$ along the rows.  We also use  $\lbrk\bm X_i\rbrk_{i\in[m]}:=[\bm X_1,\ldots,\bm X_m]\in\mathbb{R}^{s\times mt}$  to denote concatenation along the columns.
For $\bm{X} \in\mathbb{R}^{m\times d}$, we use $\mathrm{VEC}(\bm{X})$ to denote the vector in $\mathbb{R}^{md}$ constructed by concatenating the rows $\bm{X}_{i,:},1\leq i\leq m$. For $\bm{x}\in \mathbb{R}^{md}$, we use $\mathrm{MAT}(\bm{x},m)$ to denote the inverse of $\text{VEC}$, i.e. constructing a matrix $\bm{X}\in\mathbb{R}^{m\times d}$ from the vector $\bm{x}$ by consecutively slicing $\bm{x}$ into $m$ vectors in $\mathbb{R}^d$ and concatenating them along the rows.  We use $\lceil \cdot \rceil$ to denote the ceiling operation of real numbers.

A countable set $\{\phi_i\}_{i\in \mathcal{I}}$ of a real Hilbert space $\mathcal{H}$ is called a tight frame (with frame bounds equal to $1$) of $\mathcal{H}$ if and only if 
\[
\Vert f\Vert^2_\mathcal{H} = \sum_{i\in \mathcal{I}} |\langle f,\phi_i\rangle_{\mathcal{H}}|^2,\quad \forall f \in \mathcal{H},
\]
where  $\langle \cdot,\cdot\rangle_{\mathcal{H}}, \Vert \cdot \Vert_\mathcal{H}$ are the inner-product and the induced norm on $\mathcal{H}$, respectively. The condition above is equivalent to
\begin{equation}\label{eq_p_1}
	f = \sum_{i\in\mathcal{I}} \langle f,\phi_i\rangle_{\mathcal{H}}\phi_i, \quad \forall f \in \mathcal{H}.
\end{equation}
Thus, we can decompose and reconstruct any $f\in\mathcal{H}$ using $\{\phi_i\}_{i\in \mathcal{I}}$ alone. The focus of this paper is to construct tight graph framelet systems for $\mathcal{H}=L^2(\mathcal{G})$.

Under an abuse of notation, we regard $\Span(\bm{X})$ as the span of the row vectors of a matrix $\bm{X}$. We have the following key lemma concerning the characterization of the span of a matrix. 

\begin{lemma}\label{lm_1}
Let $\bm{X} := \llbracket \bm \xi_i\rrbracket_{i\in[m]}\in \mathbb{R}^{m\times n}$ be defined from the set  $\{\bm{\xi}_i\}_{i=1}^m$ of orthonormal row vectors in $\mathbb{R}^n$.  Let $\bm{A}\in \mathbb{R}^{m_1\times m}$ and $\bm{B}\in \mathbb{R}^{m_2\times m}$ be two matrices such that  $1\leq m_1< m$. 
Define $\bm{\Phi} :=\llbracket \bm{\varphi}_i\rrbracket_{i\in[m_1]} :=\bm{A} \bm{X}$ and $\bm{\Psi} := \llbracket \bm{\psi}_i\rrbracket_{i\in[m_2]}  :=\bm{B} \bm{X}$. Then the following two statements are equivalent. 

\begin{itemize}
\item[(a)] The matrices $\bm{A}$ and  $\bm{B}$ satisfy
$\bm{A}\bm{A}^\top = \bm{I}$, $\bm{B}\bm{A}^\top = \bm{0}$, and $\text{\upshape{rank}}(\bm{B})= m-m_1$. 

\item[(b)] $\Span (\bm{X}) = \Span (\bm{\Phi}) \oplus \Span (\bm{\Psi})$ and the set $\{\bm{\varphi}_i\}_{i \in[m_1]}$ is an orthonormnal basis for $\Span (\bm{\Phi})$. 
\end{itemize}
Moreover, with the additional assumption of either (a) or (b), the following statements are equivalent.
\begin{itemize}
\item[(i)] $\bm{A}^\top \bm{A} + \bm{B}^\top \bm{B} = \bm{I}$. 
\item[(ii)] $\bm{\Phi}^\top \bm{\Phi} + \bm{\Psi}^\top \bm{\Psi} = \bm{X}^\top \bm{X}$. 
\item[(iii)] $\bm{B} \bm{B}^\top \bm{B} = \bm{B}$.
\item[(iv)] $\{\bm{\psi}_i\}_{i\in[m_2]}$ is a tight frame of $\Span (\bm{\Psi})$. 
\end{itemize} 
\end{lemma}

\begin{proof} It is obvious that  (a)$\Rightarrow$(b). For (b)$\Rightarrow$(a), in view of {$\bm X\bm{X}^\top = \bm{I}$ and $\bm\Phi\bm{\Phi}^\top = \bm{I}$}, we have $\bm A \bm A^\top = \bm I$. The condition $\bm B\bm A^\top = \bm 0$ follows from  $\Span (\bm{\Psi})\perp \Span (\bm{\Phi})$.  By the isomorphism between   
$\mathbb{R}^m$ and $\Span (\bm{X}) = \Span (\bm{\Phi}) \oplus \Span (\bm{\Psi})$, we deduce that $\mathbb{R}^m=\Span(\bm A)\oplus \Span(\bm B)$, which implies the full rank condition $\mathrm{rank}(B)=m-m_1$ of $\bm B$. Hence (b)$\Rightarrow$(a). Consequently, (a)$\Leftrightarrow$(b).

Now suppose (a) holds. We next prove that (i) -- (iv) are equivalent. In fact, the equivalence between (i) and (ii) is obvious.  From \cite[Lemma 1]{li2022convolutional}, we see that $\bm{B} \bm{B}^\top \bm{B} = \bm{B}$ is equivalent to $\{\bm{B}_{i,:}\}_{i\in[m_2]}$  is a tight frame of $\Span(\bm{B})$, which is equivalent to $\{\bm{\psi}_i\}_{ i\in[m_2]}$ is a tight frame of $\Span(\bm{\Psi})$ in view of (b). Hence, (iii) and (iv) are equivalent. We only need to show (i) and (iii) are equivalent. Obviously,  (i)$\Rightarrow$(iii) by the condition $\bm B \bm A^\top = \bm 0$. Now given (iii), we see that $\bm B(\bm A^\top \bm A + \bm B^\top \bm B) = \bm B$, which implies $\bm B(\bm A^\top \bm A + \bm B^\top \bm B-\bm I) = \bm 0$. By the full rank condition of $\bm B$, we deduce (i). Therefore, (i) -- (iv) are equivalent. 
\end{proof}

Lemma~\ref{lm_1} can be extended to the following more general result, which will be useful in our later construction of graph framelet systems. 

\begin{corollary}\label{cor1}
Let $\bm X_i:=\llbracket \bm{\xi}_{i,\ell}\rrbracket_{\ell\in[r]}\in\mathbb{R}^{r\times n}$, $i\in[m]$ be matrices from row vectors $\bm\xi_{i,\ell}\in\mathbb{R}^{n}$ such that  $\Span(\bm X_i)\perp \Span(\bm X_{i'})$ and $\bm X_i \bm X_i^\top = \bm I$ for each $i,i'\in[m]$ and $i\neq i'$.  Define $\bm \Xi_\ell := \llbracket \bm{\xi}_{i,\ell}\rrbracket_{i\in[m]}\in \mathbb{R}^{m\times n}$ for each $\ell\in[r]$. 
Let  $\bm{A}\in \mathbb{R}^{m_1\times m}$ and $\bm{B}\in \mathbb{R}^{m_2\times m}$ be two matrices with $1\le m_1<m$ satisfying  $\bm{A}\bm{A}^\top = \bm{I}$, $\bm{B}\bm{A}^\top = \bm{0}$, and $\bm{B}^\top\bm{B} = \bm{I} -\bm{A}^\top\bm{A}$. Define $\bm{\Phi}_\ell :=\llbracket \bm{\varphi}_{\ell,k}\rrbracket_{k\in[m_1]} :=\bm{A} \bm\Xi_\ell$ and  $\bm{\Psi}_\ell := \llbracket \bm{\psi}_{\ell,k}\rrbracket_{k\in[m_2]} :=\bm{B} \bm \Xi_\ell$ for each $\ell\in[r]$.
Then, the following statements hold:
\begin{itemize}
\item[(1)]  $\mathbb V_{\bm X} = \mathbb V\oplus \mathbb W$ and $\mathbb V\perp \mathbb W$,
 
\item[(2)]   $\{\bm{\varphi}_{\ell,k}\}_{\ell\in[r], k\in[m_1]}$  is an orthonormal basis of  $\mathbb V$, and
 
\item[(3)]  $\{\bm{\psi}_{\ell,k}\}_{\ell\in[r], k\in[m_2]}$  is a tight frame of  $\mathbb W$,
\end{itemize} 
where 
 \begin{align}
 \mathbb V_{\bm X}&:=\Span(\bm X)  \,\,with\,\,   \bm X := \llbracket \bm X_i\rrbracket_{i\in[m]},&\\
 \mathbb V&:=\Span(\bm \Phi)\,\,\, with\,\,  \bm\Phi:=\llbracket \bm\Phi_\ell\rrbracket_{\ell \in[r]},&\\
 \mathbb W&:=\Span(\bm\Psi) \,\,\, with\,\,   \bm\Psi:=\llbracket \bm\Psi_\ell\rrbracket_{\ell\in[r]}.&
 \end{align}
\end{corollary}

\begin{proof}
Note that the matrices  $\bm \Xi_1,\ldots, \bm \Xi_r$  are obtained from the simple rearrangement of the rows of the matrices $\bm X_1,\ldots, \bm X_m$, that is, there exists a permutation matrix $\bm E$ of size $mr\times mr$ such that 
\begin{equation}\label{permu:E}
\bm \Xi = \bm E \bm X,
\end{equation}
where $\bm \Xi := \llbracket \bm \Xi_\ell\rrbracket_{\ell\in[r]}$. 
Moreover, the matrix  $\bm X \in \mathbb{R}^{mr\times n}$ satisfies $\bm X\bm X^\top = \bm I$.  Let 
\begin{equation}\label{A:B}
\begin{aligned}
\left\{
\begin{matrix}
\mathcal A := \mathrm{diag}(\bm A,\ldots, \bm A) \bm E\in\mathbb{R}^{m_1r\times mr},\\
\mathcal B := \mathrm{diag}(\bm B,\ldots, \bm B)\bm E\in \mathbb{R}^{m_2r\times mr}. 
\end{matrix}
\right.
\end{aligned}
\end{equation}
Then, we have $\mathcal A\mathcal A^\top = \bm I$, $\mathcal B \mathcal A^\top = \bm 0$, and $\mathcal B^\top \mathcal B = \bm  I - \mathcal A^\top\mathcal A$. Moreover, $\bm \Phi = \mathcal A \bm X$ and $\bm \Psi = \mathcal B \bm X$.
The conclusion then directly follows from applying Lemma~\ref{lm_1}. 
\end{proof}

%
\begin{remark}
{\rm 
We call the matrices $\bm A, \bm B$ in Corollary~\ref{cor1} as \emph{filters or filter matrices},  and the matrices $\bm X_i, i\in[m]$ as \emph{subspace matrices}. 
We formulate Corollary \ref{cor1} as an algorithm in Algorithm \ref{alg:Algo1}. We highlight that the key of Corollary~\ref{cor1} is that each subspace $\Span(\bm X_i)$ has the same dimension $r$ for all $i\in[m]$. Such a condition will play a crucial role in our later construction of tight graph framelet systems. 
}
\end{remark}
\begin{algorithm}[htb!]\
		\caption{Subspace Decomposition}			           \label{alg:Algo1}
		\begin{algorithmic}[1]
			\State {\bfseries Input:} Matrices  $\bm X_i = \llbracket \bm{\xi}_{i,j}\rrbracket_{j\in[r]}\in\mathbb{R}^{r\times n}$ , $i\in[m]$, $\bm{A}\in \mathbb{R}^{m_1\times m}$, $\bm{B}\in \mathbb{R}^{m_2\times m}$  with $1\leq m_1< m$.
			\State $\bm{\Xi}_{\ell}:=\text{\upshape{CONCAT}}\left(\{\bm{\xi}_{i,\ell}\}_{i=1}^m\right),\, \ell\in [r]$.
			\State $\bm{\Phi}:= \text{\upshape{CONCAT}}\left(\left\{\bm{A}\bm{\Xi}_{\ell}\right\}_{\ell=1}^r\right)$.
			\State $\bm{\Psi}:=\text{\upshape{CONCAT}}\left(\left\{\bm{B}\bm{\Xi}_{\ell}\right\}_{\ell=1}^r\right)$.
			\State {\bfseries Output:} $\bm\Phi$ and $\bm\Psi$.
		\end{algorithmic}
\end{algorithm}
%

\subsection{Construction of $\mathcal{V}$-Framelet Systems}
We next provide the details for the construction of framelet systems based on a partition tree for a vertex set $\mathcal{V}$, which are the fundamental structures for our construction of graph framelet systems. 

\begin{definition}\label{def1}
Let   $J\in \mathbb{N}$ and $\mathcal{V}=\{v_1,\dots,v_n\}$ be a set of $n$ vertices.
A \emph{partition tree} $\mathscr{T}_J({\mathcal{V}})$ of $\mathcal{V}$ with $J+1$ levels is a rooted tree such that
\begin{itemize}

\item[(a)] {The root node $p_{0,1}$ is associated with $\mathcal{S}_{0,1}:=\mathcal{V}$.}

\item[(b)]{Each leaf $p_{J,k},k\in [n_J]$ with $n_J:=n$ is associated with the singleton $\mathcal{S}_{J,k}:=\{v_i\}$. The path from $p_{0,1}$ to each $p_{J,k}$ contains exactly $J$ edges.}

\item[(c)]Each tree node $p_{j,k},k \in [n_j]$ on the $j \in [J-1]$ level (i.e. the path from $p_{0,1}$ to $p_{j,k}$ contains exactly $j$ edges) is associated with a set $\mathcal{S}_{j,k}\subset \mathcal{V}$ of vertices such that 
\begin{itemize}
\item[(i)] $\cup_{k=1}^{n_j} \mathcal{S}_{j,k} = \mathcal{V}$, $\mathcal{S}_{j,k_1}\cap\mathcal{S}_{j,k_2} = \varnothing$ for $k_1\neq k_2$, $1\le k_1,k_2\le n_j$,  and
\item[(ii)] $\cup_{k'\in \mathcal{C}_{j,k}} \mathcal{S}_{j+1,k'} = \mathcal{S}_{j,k}$ with $\mathcal{S}_{j+1,k'}\subsetneq \mathcal{S}_{j,k}$,
\end{itemize}
where $n_j<n$ is the number of tree nodes on level $j$ and  $\mathcal{C}_{j,k}\subseteq [n_{j+1}]$ is the index set of children nodes of $p_{j,k}$.
\end{itemize}
\end{definition}

\begin{figure}[htpb!]
	\centering
	\includegraphics[width=0.8\linewidth]{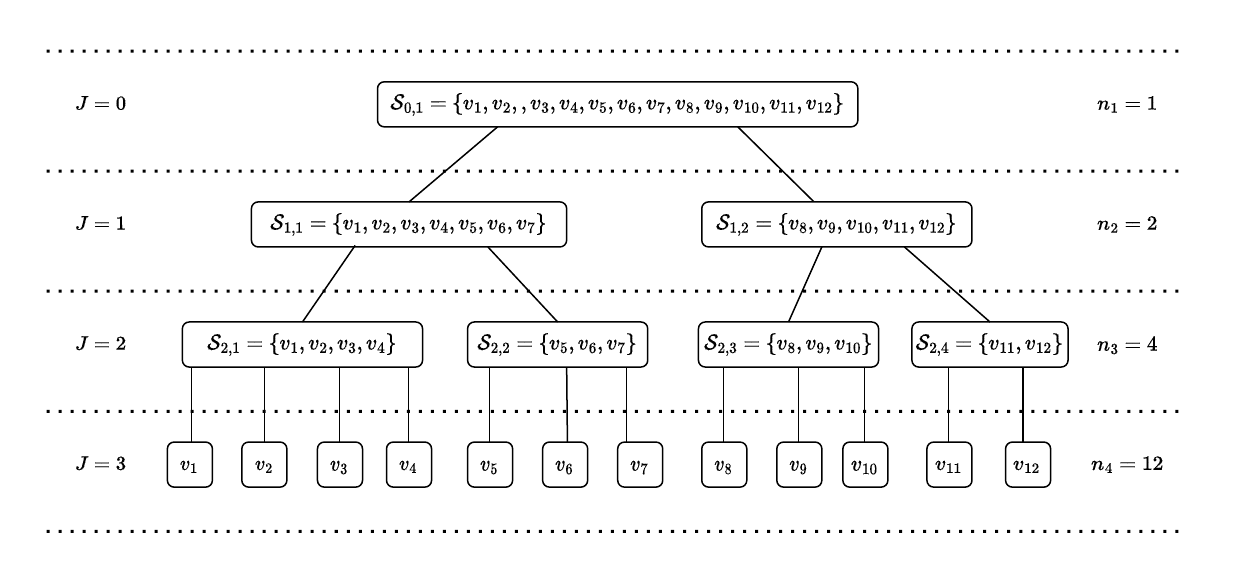}
	\caption{A partition tree $\mathscr{T}_J(\mathcal{V})$ for a vertex set $\mathcal{V}$ of 12 vertices and with $J=3$ (4 levels).}
	\label{fig:tree}
\end{figure}

See Figure \ref{fig:tree} for an illustration. In short, each level in the partition tree is associated with a partition on $\mathcal{V}$, and the partitions in the lower levels are formed by merging clusters in the higher levels. Note that by the condition in item (c), each non-leaf node has at least $2$ children, i.e., $|\mathcal{C}_{j,k}| > 1$. 

In order to construct graph framelet systems analog to the classical wavelet/framelet systems with the multiscale structure, based on a partition tree $\mathscr{T}_J({\mathcal{V}})$,  we associate each level $j$ with a linear subspace $\mathbb V_j$ such that $\mathbb V_0\subsetneq \mathbb V_1 \subsetneq \dots \subsetneq \mathbb V_J$. Following Mallat's idea in Multi-resolution analysis (MRA) \cite{mallat1989theory}, if we have 
\begin{equation}\label{eq7}
\mathbb V_j = \mathbb V_{j-1}\oplus \mathbb W_{j-1},\quad \mathbb V_{j-1}\perp \mathbb W_{j-1},\quad j \in [J], 
\end{equation}
then by doing the decomposition $J-1$ times, we have 
\begin{equation}
\mathbb V_J =\mathbb  V_0\oplus\mathbb  W_0 \oplus\mathbb  W_1\oplus \dots \oplus \mathbb W_{J-1} \label{eq3_2}
\end{equation} 
and that $\mathbb V_0, \mathbb W_0,\mathbb W_1,\dots, \mathbb W_{J-1}$ are mutually orthogonal. Therefore, to construct a tight frame on $\mathbb V_J$, it is sufficient to construct tight frames for each $\mathbb V_0, \mathbb W_0,\mathbb W_1, \dots, \mathbb W_{J-1}$ and take the union.

The partition tree $\mathscr{T}_J({\mathcal{V}})$ suggests that each $\mathbb V_j$ and $\mathbb W_j$ should be further decomposed such that
\begin{equation}
	\mathbb V_j := \oplus_{k=1}^{n_j}\mathbb  V_{j,k},\quad \mathbb W_j := \oplus_{k=1}^{n_j} \mathbb W_{j,k}\label{eq3_3},
	\end{equation}
where each element in $\mathbb V_{j,k}$ or $\mathbb W_{j,k}$ is supported on $\mathcal{S}_{j,k}$ and thus $\mathbb V_{j,k},  k \in[n_j]$ (or $\mathbb W_{j,k},k \in [n_j]$) are mutually orthogonal.
Furthermore, the nested structure of $\mathscr{T}_J({\mathcal{V}})$ further suggests that
\begin{equation}
\begin{gathered}\label{eq3_4}
\oplus_{k'\in \mathcal{C}_{j,k}} \mathbb V_{j+1,k'} = \mathbb V_{j,k}\oplus \mathbb W_{j,k},\quad \mathbb V_{j,k}\perp \mathbb W_{j,k}
\end{gathered}
\end{equation}
for $j \in \{0,\dots,J-1\}$ and $k \in [n_j]$. 

Assuming that  each $\mathbb V_{J,k}$ is supported on $\mathcal{S}_{J,k}=\{v_k\}$ at the finest level $J$, 
we can define $\mathbb V_J := \Span\{\bm e_k\setsep k\in[n]\}=\oplus_{k=1}^n \mathbb V_{J,k}$, where $\mathbb V_{J,k}:=\Span(\{\bm  e_k\})$ is simply the one-dimensional linear space associated with the vertex $v_k$ and generated by the $k$-th canonical basis vector $\bm e_k:=[0,\ldots, 0,1,0\ldots,0]\in\mathbb{R}^n$. Note that by the parent-children relation, we can rewrite $\mathbb V_J$ as
$\mathbb V_J  =\oplus_{k\in [n_{J-1}]}(\oplus_{k'\in \mathcal C_{J-1,k}}\mathbb V_{J,k'})$.
{It is straightforward to see that \eqref{eq3_3} and \eqref{eq3_4} implies \eqref{eq7}}, which eventually implies (\ref{eq3_2}) since we can decompose 
$\oplus_{k'\in \mathcal C_{J-1,k}}\mathbb V_{J,k'} $ as 
$\oplus_{k'\in \mathcal{C}_{J-1,k}}\mathbb V_{J,k'}  = \mathbb V_{J-1,k}\oplus \mathbb W_{J-1,k}$
for each $k\in[n_{J-1}]$, and  the procedure can continue from bottom $j=J$ to top $j=0$. Hence, it remains to define a general procedure of decomposition for each non-leaf node in $\mathscr{T}_J({\mathcal{V}})$ such that (\ref{eq3_4}) is satisfied. 
Our next result shows that we indeed can define such a general (bottom-up) procedure and obtain a tight framelet system for $\mathbb V_J$ 
based on Cororllary~\ref{cor1} and the partition tree $\mathscr{T}_J(\mathcal V)$. 

We first define the $\mathcal V$-framelet system based on the partition tree $\mathscr{T}_J(\mathcal V)$.
\begin{definition}\label{def2}
Let $\mathscr{T}_J({\mathcal{V}})$ be a partition tree as in Definition~\ref{def1}. Let $r_j$ be an integer such that  $r_j \in [\min(|\mathcal{C}_{j,1}|,\dots,|\mathcal{C}_{j,n_j})|-1]$ and  $\bm{A}^{[j,k]}\in \mathbb{R}^{r_j\times |\mathcal{C}_{j,k}|}$ and $\bm{B}^{[j,k]}\in \mathbb{R}^{m_{j,k}\times |\mathcal{C}_{j,k}|}$  be two filter matrices associated with the tree node $p_{j,k}$ for $j=0,\ldots, J-1$ and $k\in[n_j]$. 
\begin{itemize}
\item[(1)] {\rm (Initialization)}  Define  $\mathbb V_{J,k}:=\Span(\{\bm  e_k\})$ and $\bm{\Phi}^{[J,k]}:=\bm e_k$ for  $k\in [n]$. Let $\mathbb V_J :=\oplus_{k\in[n]} \mathbb V_{J,k}$ and $r_J:=1$.

\item[(2)] {\rm (Bottom-up Procedure)} Recursively define at each level $j$ from $J-1$ to $0$: 

\begin{itemize}
\item[(a)] For each  tree node $p_{j,k}$, obtain $\mathbb V_{j,k}:=\Span(\bm\Phi^{[j,k]})$ 
and $\mathbb W_{j,k}=\Span(\bm\Psi^{[j,k]})$ by Corollary~\ref{cor1} with subspace matrices  $\{\bm X_{k'}:= \bm\Phi^{[j+1,k']} \setsep k'\in \mathcal C_{j,k}\}$  and the two filter matrices $\bm A^{[j,k]}$ and $\bm B^{[j,k]}$. More precisely,  let $\bm X:=\llbracket \bm X_{k'}\rrbracket_{k'\in\mathcal C_{j,k}}$. Then we have 
\begin{equation}
\label{AB:jk}
\bm \Phi^{[j,k]} = \mathcal A^{[j,k]} \bm X,\quad 
\bm \Psi^{[j,k]} = \mathcal B^{[j,k]} \bm X,
\end{equation}
where $\mathcal A^{[j,k]}, \mathcal B^{[j,k]}$ are  matrices corresponding to \eqref{A:B} in Corollary~\ref{cor1}.

\item[(b)] Define $\mathbb V_j:=\Span(\bm\Phi_j)$ with $\bm\Phi_j:=\llbracket \bm \Phi^{[j,k]}\rrbracket_{k\in[n_j]}$ and $\mathbb W_j:=\Span(\bm\Psi_j)$ with $\bm\Psi_j:=\llbracket\bm \Psi^{[j,k]}\rrbracket_{k\in[n_j]}$. 
\end{itemize}

\item[(3)]{\rm (Finalization)} For each $J_0=0,\ldots, J$,  define the \emph{$\mathcal V$-framelet system}  associate with the partition tree $\mathscr{T}_J(\mathcal V)$ and determined by the filter matrices as 
\begin{equation}\label{def:vfmt}
\begin{aligned}
\mathcal{F}_{J_0}^J(\mathcal{V}):=&\mathcal F_{J_0}(\{(\bm A^{[j,k]}, \bm B^{[j,k]})\setsep k\in[n_j] \}_{j=J_0}^{J-1})
\\:=&\bm\Phi_{J_0} \cup\bm \Psi_{J_0}\cup\cdots\cup\bm \Psi_{J-1}.
\end{aligned}
\end{equation}
\end{itemize}
\end{definition}

Now we are ready to state the main theorem for the construction of a general tight $\mathcal V$-framelet system for $\mathbb V_J$. 

\begin{theorem}\label{thm_1}
Adopt the notations in Definition \ref{def2}. Assume that the filter matrices $\bm{A}^{[j,k]}$ and $\bm{B}^{[j,k]}$
satisfy 
\begin{align}
	\bm{A}^{[j,k]}({\bm{A}^{[j,k]}})^\top &= \bm{I},\label{eq3:thm1a}\\
	\bm{B}^{[j,k]}({\bm{A}^{[j,k]}})^\top &= \bm{0},\label{eq3:thm1b}\\
	({\bm{B}^{[j,k]}})^\top\bm{B}^{[j,k]}& = \bm{I} -({\bm{A}^{[j,k]}})^\top\bm{A}^{[j,k]}, \label{eq3:thm1c}
\end{align}	
for $j\in \{0,\dots,J-1\}$ and $k\in [n_j]$.
Then the following statements hold.
\begin{itemize}
\item[\rm{(i)}]
$\mathbb V_j = \oplus_{k=1}^{n_j} \mathbb V_{j,k}$, $\mathbb W_j = \oplus_{k=1}^{n_j} \mathbb W_{j,k}$ for $j=0,\ldots, J-1$, and $\mathbb V_j = \mathbb V_{j-1}\oplus \mathbb W_{j-1}, \mathbb V_{j-1} \perp \mathbb W_{j-1}$ for  $j\in[J]$.

\item[\rm{(ii)}]	  $\mathbb V_J=\mathbb  V_j\oplus\mathbb  W_j \oplus \mathbb W_{j+1}\oplus \dots \oplus \mathbb  W_{J-1}$ for $j=0,\ldots, J-1$.  In particular, $\mathbb V_J=\mathbb  V_0\oplus\mathbb  W_0 \oplus \mathbb W_{1}\oplus \dots \oplus \mathbb  W_{J-1}$.

\item[\rm{(iii)}]	 $\mathcal F_{J_0}^J(\mathcal V)$  is a tight frame for $\mathbb V_J$ for all $J_0 = 0,\ldots, J-1$.
\end{itemize}
\end{theorem}

\begin{proof}
We apply mathematical induction on $j$ (decreasingly) to prove that for each $j\in{0,\dots,J-1}$, all the subspaces $\mathbb V_{j+1,k'}=\Span(\bm\Phi^{[j+1,k']})$ in the previous level have the same dimension and conditions in Corrolary~\ref{cor1} are satisfied with $\bm X_k':=\bm\Phi^{[j+1,k']}, k'\in \mathcal{C}_{j,k}$. Hence, by the results of Corollary~\ref{cor1}, we can obtain the conclusions of items (i)--(iii) of the theorem. 

\begin{itemize}
\item[(a)] 
Base case: By definition, each $\mathbb V_{J,k},k\in[n]$ is spaned by $ \bm\Phi^{[j+1,k]}=\bm e_{k}\in \mathbb{R}^{r_J\times n}$, which is $r_J$-dimensional ($r_J=1$). Note that $\mathbb V_J = \oplus_{k\in[n_{J-1}]} \oplus_{k'\in\mathcal C_{J-1,k}}\mathbb V_{J,k'}$.

\item[(b)] Induction hypothesis:   Assume that at level $j+1$ with $0\le j\le J-1$, each $\mathbb V_{j+1,k}=\Span(\bm\Phi^{[j+1,k]})$  has the same dimension $R_{j+1}:=\prod_{i=j+1}^J r_{i}$ for all $k\in [n_{j+1}]$, where each $\bm\Phi^{[j+1,k]}$ is formed by othonormal row vectors supported on $\mathcal{S}_{j+1,k}$. 

\item[(c)] Induction step:
We next prove that each $\mathbb V_{j,k}=\Span(\bm\Phi^{[j,k]})$ also has the  same dimension $R_j:=\prod_{i=j}^J r_{i}$ for all $k\in[n_j]$ with each $\bm\Phi^{[j,k]}$ being formed by othonormal row vectors. 
In fact,  fixed the node $p_{j,k}$, it has its children nodes $p_{j+1,k'},  k'\in \mathcal C_{j,k}$.  Consider $\bm X_{k'} :=\bm\Phi^{[j+1,k']}\in \mathbb{R}^{R_{j+1}\times n}$, $k'\in\mathcal{C}_{j,k}$. By \eqref{eq3:thm1a} -- \eqref{eq3:thm1c}, the conditions of Corollary~\ref{cor1}  are satisfied with these subspace matrices $\bm X_{k'}$ and the two filter matrices $\bm A^{[j,k]}, \bm B^{[j,k]}$. Thus its conclusions can be obtained. 
That is, $\mathbb V_{\bm X} := \oplus_{k'\in \mathcal C_{j,k}} \mathbb V_{j+1,k'} = \mathbb V_{j,k} \oplus \mathbb W_{j,k},  \mathbb V_{j,k} \perp \mathbb W_{j,k}$, and the rows of $\bm\Phi^{[j,k]}$ form an orthonormal basis for $\mathbb V_{j,k}$. Moreover, the rows of $\bm\Psi^{[j,k]}$ form a tight frame for $\mathbb W_{j,k}$. Note that the dimension of $\mathbb V_{j,k}$ satisfies $\mathrm{dim}(\mathbb V_{j,k}) \equiv r_{j}R_{j+1} = \prod_{i=j}^J r_i$ for all $k\in [n_{j}]$. 

\item[(d)] Conclusion: Therefore, for each $j=0,\ldots, J$, the space $\mathbb V_{j,k}=\Span(\bm\Phi^{[j,k]})$  has the same dimension $R_{j}$ for all $k\in [n_{j}]$ with each $\bm\Phi^{[j,k]}$ being formed by orthonormal row vectors.

\end{itemize}

Consequently, item (i) holds and it implies item (ii). Item (iii) follows from item (ii) and the result that the rows of $\bm \Psi^{[j,k]}$ form a tight frame for $\mathbb W_{j,k}$. 
\end{proof}

The construction of $\mathcal V$-framelet system in Theorem \ref{thm_1} is formulated in Algorithm \ref{alg:Algo2}. Before we proceed, we have the following remarks. 

\begin{algorithm}[!t]\
		\caption{Generation of the $\mathcal V$-Framelet System }			           \label{alg:Algo2}
		\begin{algorithmic}[1]
			\State {\bfseries Input:} A partition tree $\mathscr{T}_J({\mathcal{V}})$, a set of postive integers $\{r_j\}_{j=1}^J$, matrices $\bm{A}^{[j,k]}\in \mathbb{R}^{r_j\times |\mathcal{C}_{j,k}|},\bm{B}^{[j,k]}\in \mathbb{R}^{m_{j,k}\times |\mathcal{C}_{j,k}|},k\in [n_j],j\in\{0,\dots,J-1\}$.
			\State $\bm{\Phi}^{[J,k]}:=\bm e_k, k\in [n]$.
	\For{$j=J-1$ {\bfseries to} $0$}
	\For{$k=1$ {\bfseries to} $n_j$}
\State $\bm{\Xi}_{\ell}:=\text{\upshape{CONCAT}}(\{\bm{\Phi}^{[j+1,k']}_{\ell,:}\}_{k'\in \mathcal{C}_{j,k}}),\, \ell\in [r_j]$.
			\State $\bm{\Phi}^{[j,k]}:= \text{\upshape{CONCAT}}(\{\bm{A}^{[j,k]}\bm{\Xi}_{\ell}\}_{\ell=1}^{r_j})$.
			\State $\bm{\Psi}^{[j,k]}:=\text{\upshape{CONCAT}}(\{\bm{B}^{[j,k]}\bm{\Xi}_{\ell}\}_{\ell=1}^{r_j})$.
			

\EndFor
	\State $\bm{\Phi}_j := \cup_{k=1}^{n_j} \bm{\Phi}^{[j,k]},\quad \bm{\Psi}_j := \cup_{k=1}^{n_j} \bm{\Psi}^{[j,k]}$.

\EndFor
			\State {\bfseries Output:} $\{\bm{\Phi}_0\}\cup \{\bm{\Psi}_j\}_{j=0}^{J-1}$.
		\end{algorithmic}
\end{algorithm}

\begin{remark}\label{rem1}
{\rm
Note that the construction of $\mathcal V$-framelet systems only involves the vertex set $\mathcal{V}$ and the partition tree $\mathscr{T}_J(\mathcal V)$.  No graph structure, such as the edge set $\mathcal E$ or the adjacency matrix, is involved. Such construction based on the hierarchical partitions of a vertex set has the benefit of providing a mathematical generality. We discuss the graph framelet systems in the next subsection that utilize both the vertex set $\mathcal V$ and the edge set $\mathcal E$. 
}
\end{remark}

\begin{remark}
{\rm
As in classical wavelet/framelet theory, we called the elements of $\bm{\Phi}_j$ \textbf{scaling functions} and the elements of $\bm{\Psi}_j$ \textbf{framelets}. The framelets have \textbf{localized supports}, i.e. framelets in $\mathbb W_{j,k}$ are supported on $\mathcal{S}_{j,k}$. 
}
\end{remark}

\begin{remark}
{\rm
The dimension of $\mathbb V_j$ is $R_j = \Pi_{i=j}^J r_i$, which can be controlled by varying the $r_i$.  As mentioned in the introduction,  the orthogonality to $\mathbb V_j$ can be used to define vanishing moments of the framelet system. Note that the $\mathcal V$-framelet system $\mathcal{F}_{J_0}^J(\mathcal V)$ as defined in \eqref{def:vfmt} has its framelets all from $\mathbb W_{j}$ for $j\ge J_0$ and hence all orthogonal to $\mathbb V_{J_0}$. We call the  such a framelet system $\mathcal{F}_{J_0}^J(\mathcal V)$ has \textbf{generalized} $R_{J_0}$ \textbf{vanishing moments}. 
}
\end{remark}



\begin{remark}
{\rm The condition in (\ref{eq3:thm1c}) can be considered as a generalized version of the \emph{unitary extension principle} in classic wavelet/framelet theories \cite{ron1997affine}, which requires  
the filter bank to form a partition of unity. By analogy, we can regard $\bm{A}^{[j,k]},\bm{B}^{[j,k]}$ as \emph{filters} and their union as a \emph{filter bank}.}
\end{remark}

\begin{remark}
{\rm 
In \cite{2023arXiv230604265L}, Haar-type tight  framelet system is obtained by defining $\bm{A}^{[j,k]}:=\frac{1}{\sqrt{|\mathcal{C}_{j,k}|}}\begin{bmatrix}
	1,\dots,1
\end{bmatrix}
\in \mathbb{R}^{1\times |\mathcal{C}_{j,k}|}$
and
$\bm{B}^{[j,k]}\in \mathbb{R}^{\frac{|\mathcal{C}_{j,k}|(|\mathcal{C}_{j,k}|-1)}{2}\times |\mathcal{C}_{j,k}|}$ whose rows 
$\bm{B}^{[j,k]}_{m,:} :=\bm{b}^{m}$, $1\leq m \leq \frac{|\mathcal{C}_{j,k}|(|\mathcal{C}_{j,k}|-1)}{2}$, 
are 
given by $\bm{b}^{m} \in \mathbb{R}^{1\times |\mathcal{C}_{j,k}|}$ with its $i$th entry being defined by 
\[
\bm{b}^{m}_{i}:=
 \left\lbrace
 \begin{aligned}
 \frac{1}{ {|\mathcal{C}_{j,k}|}} \quad &i=s,\\ 
 \frac{-1}{\sqrt{|\mathcal{C}_{j,k}|}}\quad  &i=t,\\ 
 0\quad &\text{otherwise,}
\end{aligned}
\right.
\]
where $m$ is uniquely determined by the integers  $s,t$ with $1\le s<t\le |\mathcal C_{j,k}|$ as $m:=m(s,t,|\mathcal{C}_{j,k}|):=\frac{(2|\mathcal{C}_{j,k}|-s)(s-1)}{2}+s-t$.
It is shown that (\ref{eq3:thm1c}) in Theorem~\ref{thm_1} is satisfied. In summary, the rows of $\bm{B}^{[j,k]}$ is constructed by enumerating pairs of indices in $\mathcal{C}_{j,k}$ and assigning the same constant with different signs, which resemble classic Haar wavelets/filters. Our general construction includes \cite{2023arXiv230604265L} as a special case.
}
\end{remark}

\subsection{Construction of $\mathcal G$-Framelet Systems}\label{sec:gif}
As we point out in Remark~\ref{rem1}, the $\mathcal{V}$-framelet systems do not involve the other graph structure such as the edge set $\mathcal E$. However,  they do depend on the partition tree $\mathscr{T}_J(\mathcal{V})$ and how to obtain such a partition tree is closely related to the graph $\mathcal G=(\mathcal V,\mathcal E, \bm W)$. 
We next discuss how we can obtain the partition tree from a given graph $\mathcal G$ and define the so-called graph-involved $\mathcal G$-framelet systems.

We first discuss the realization of the nested structure of a partition tree.  It is known that edges in graphs intuitively represent a notion of proximity among nodes. Thus, to represent scales (levels) on graphs analog to the Euclidean domains, a common practice is to apply a series of clustering (coarsing) on graphs. In detail, given $\mathcal{G} = (\mathcal{V},\mathcal{E},\bm{W})$ with $\mathcal V:=\{v_1,\ldots, v_n\}$, the clusters on $\mathcal{V}$ are formed through certain algorithms, e.g., $K$-means clustering, based on $\mathcal{E}$ such that connected nodes are more likely to be in the same clusters. Assuming that the clustering algorithm on $\mathcal G$ resulted in  $n'$ clusters, denoted as $\mathcal{V}' = \{v_1',\ldots, v_{n'}'\}$, where each $v_i' =\{v_{i_1},\ldots, v_{i_{n_i}}\}\subseteq \mathcal V$, $v_i'\cap v_{j}'=\varnothing$  for $i\neq j$, and $\cup_{i=1}^{n'} v_i' = \mathcal V$. Then a  graph $\mathcal{G}':=(\mathcal{V}',\mathcal{E}', \bm{W}')$ can be formed from these clusters through the definition of its adjacency matrix $\bm W'=(w_{ij}')_{1\le i,j\le n'}$ as
\[
	w_{ij}' :=\sum_{p\in v_i'}\sum_{q\in v_j'} w_{pq}, \quad i,j = 1,\ldots, n',
\]
where  $w_{ij}'$ is the weight between $v_i',v_j'\in\mathcal V'$ while  $ w_{pq}$ is the original weight between vertices $p,q\in\mathcal V$. 

In summary, the nodes of $\mathcal{G}'$ represent clusters and the edge weight on $\mathcal{G}'$ is determined by summing of the edge weight among nodes of two clusters. We called $\mathcal{G}'$ a \emph{coarse-grained graph} of $\mathcal{G}$ \cite{zheng2022decimated}. Based on the coarse-grained graph, we can give the definition of multi-graph partition trees.
\begin{definition}
A \emph{multi-graph partition tree} $\mathscr{T}_J({\mathcal{G}})$ of $\mathcal{G}=(\mathcal V,\mathcal E,\bm W)$ with $J+1,J\in \mathbb{N}$ levels is a partition tree $\mathscr{T}_J({\mathcal{V}})$ as defined in Definition~\ref{def1} such that each  $j\in \{0,\dots,J\}$ is associated with a coarse-grained graph $\mathcal{G}^j=(\mathcal V^j, \mathcal E^j)$ of $\mathcal G$ and $\mathcal V^j =\{\mathcal S_{j,k}\setsep k\in[n_j]\}$. In particular, $\mathcal G^J\equiv \mathcal G$, where we consider a vertex in $\mathcal G^J$ as a cluster of singleton. 
\end{definition}

{An intuitive and equivalent interpretation of the definition above is that a multi-graph partition tree is a partition tree by generated successively forming coarse-grained graphs $\mathcal{G}^{J-1},\mathcal{G}^{J-1}\dots,\mathcal{G}^0$ such that $\mathcal{G}^{j}$ is a coarse-grained graph of $\mathcal{G}^{j+1}, j\in [J-1]$.} Since $\mathscr{T}_J(\mathcal G)$ is associated with a partition tree $\mathscr{T}_J(\mathcal V)$, we can construct $\mathcal V$-framelet systems as in Definition~\ref{def2}. We have the following definition.

\begin{definition}\label{def_3}
Let $\mathscr{T}_J(\mathcal G)$ be a multi-graph partition tree associated with a graph $\mathcal G=(\mathcal V, \mathcal E, \bm W)$. Let \{$r_j$, $\bm{A}^{[j,k]}$, $\bm{B}^{[j,k]}\setsep j=0,\ldots,J-1; k\in[n_j]\}$ be the set of integers and filter matrices satisfying the assumptions of Theorem \ref{thm_1}. Then for  $J_0=0,\ldots, J-1$, the $\mathcal V$-framelet system $\mathcal F_{J_0}^J(\mathcal V)$ as in \eqref{def:vfmt} is a tight frame for $\mathbb V_J=\Span\{\bm e_k\setsep k\in[n]\}=L_2(\mathcal G)$, which we call  a \emph{(graph-involved) $\mathcal G$-framelet (GIF) system}. In such a case, we simply denote  $\mathcal F_{J_0}^J(\mathcal G):=\mathcal F_{J_0}^J(\mathcal V)$   to indicate the role played by $\mathcal G$. In particular, we define  for the special case $J_0=0$, the GIF system $\mathcal F_0^J(\mathcal G)$ as 
\begin{equation}\label{def:GIF}
\mathsf{GIF}_J(\mathcal{G}):=\mathcal F_0^{J}(\mathcal G)=\{\bm\Phi_0,\bm\Psi_0,\cdots,\bm \Psi_{J-1}\}.
\end{equation} 
\end{definition}

\begin{remark}
{\rm When $r_j = 1$ for all $j = \{0,\dots,J-1\}$, the filter matrix $\bm{A}^{[j,k]}$ consists of the first eigenvector of $\bm{L}^{[j,k]}$, and the rows of $\bm{B}^{[j,k]}$ are the remaining eigenvectors of $\bm{L}^{[j,k]}$, where $\bm{L}^{[j,k]}$ is the graph Laplacian matrix of the subgraph of $\mathcal{G}^{j+1}$ {induced by the nodes $\{\mathcal S_{j+1,k'}\setsep k'\in \mathcal{C}_{j,k}\}$}, then the $\mathcal G$-framelet system $\mathsf{GIF}_J(\mathcal{G})$ is an orthonormal basis, which corresponds to the case $p=2$ in \cite{tremblay2016subgraph}.
}
\end{remark}


\section{Graph Framelet Transforms}\label{sec:gmt}
In this section, we focus on the fast transform algorithms based on the $\mathcal G$-framelet systems as well as their computational complexity.

\subsection{Graph Framelet Transforms}
Similar to the discrete wavelet/framelet transforms, the iterative nature of the tight graph framelet systems $\mathcal{F}_{J_0}^J(\mathcal{G})$ for $J_0=0,\ldots, J-1$ can be used to derive the graph framelet transforms for graph signals $\bm f\in L_2(\mathcal G)$.  In fact, given such an $\bm f$, which is a row vector of size $n$, by the tightness of the system $\mathcal{F}_{J_0}^J(\mathcal{G})$, it holds for $J_0=0,\ldots, J-1$ that
\begin{equation}\label{tight:f}
\bm f = \sum_{k\in [n_{J_0}]} \bm c^{[J_0,k]} \bm\Phi^{[J_0,k]}+\sum_{j=J_0}^{J-1}\sum_{k\in [n_j]}\bm d^{[j,k]} \bm\Psi^{[j,k]},
\end{equation}
where 
\begin{equation}\label{coefs}
\bm c^{[j,k]}:= \langle \bm f, \bm \Phi^{[j,k]}\rangle, \quad 
\bm d^{[j,k]}:=  \langle \bm f, \bm \Psi^{[j,k]}\rangle,
\end{equation}
and the inner product is defined by
\[
\langle \bm A, \bm B\rangle:=\bm A\bm B^\top, \quad \bm A\in \mathbb{R}^{r\times m}, \bm B\in\mathbb{R}^{s\times m}.
\] 
The \emph{graph framelet transforms} consists of the analysis and the synthesis phases, where the analysis step is to decompose a given signal $\bm f$ into its framelet coefficient sequences: $\bm c^{[J_0,k]}, \bm d^{[j,k]}$,  $j=J_0,\ldots, J-1, k\in[n_j]$, and the synthesis step is to reconstruct a signal $\bm f$ from such a given framelet coefficient sequences. Recall that $\llbracket\cdot\rrbracket$ and $\lbrk\cdot\rbrk$ are row-wise and column-wise concatenation, respectively. The following result discusses the one-level decomposition and reconstruction of the graph framelet transforms. 

\begin{corollary}\label{cor_1}
Let $\bm f\in L_2(\mathcal G)$ be a graph signal and  $\bm{c}^{[j,k]},\bm{d}^{[j,k]}$ denote the vector coefficients with respect to  $\bm{\Phi}^{[j,k]}, \bm{\Psi}^{[j,k]}$ as defined in \eqref{coefs} from the tight graph framelet system $\mathsf{GIF}_J(\mathcal G)$. Then
\begin{equation}\label{eq:dec1}
\begin{aligned}
\left\{
\begin{matrix}
\bm{c}^{[j,k]} & = \langle \lbrk \bm c^{[j+1,k']}\rbrk_{k'\in \mathcal C_{j,k}}, \mathcal A^{[j,k]}\rangle, \\
\bm{d}^{[j,k]} & = \langle \lbrk\bm c^{[j+1,k']}\rbrk_{k'\in \mathcal C_{j,k}}, \mathcal B^{[j,k]}\rangle, \\
\end{matrix}
\right.
\end{aligned}
\end{equation}
and
\begin{equation}\label{eq:rec1}
(\bm{c}^{[j+1,k']})_{k'\in \mathcal C_{j,k}} = \bm c^{[j,k]}\mathcal A^{[j,k]} + \bm d^{[j,k]}\mathcal B^{[j,k]},\\
\end{equation}
where $\mathcal A^{[j,k]}, \mathcal B^{[j,k]}$ are those matrices as in \eqref{AB:jk}  satisfying $(\mathcal A^{[j,k]})^\top \mathcal A^{[j,k]}+(\mathcal B^{[j,k]})^\top \mathcal B^{[j,k]} = \bm I$. 
\end{corollary}

\begin{proof}
Note that
\[
\begin{aligned}
\bm{c}^{[j,k]} & = \langle \bm  f, \bm \Phi^{[j,k]}\rangle 
=\langle \bm  f, \mathcal A^{[j,k]} \llbracket \bm \Phi^{[j+1,k']}\rrbracket_{k'\in \mathcal C_{j,k}}\rangle \\
&=\bm  f\llbracket\bm \Phi^{[j+1,k']}\rrbracket_{k'\in \mathcal C_{j,k}}^\top (\mathcal A^{[j,k]})^\top \\
&=\lbrk \langle \bm  f,\bm \Phi^{[j+1,k']}\rangle\rbrk_{k'\in \mathcal C_{j,k}} (\mathcal A^{[j,k]})^\top \\
&=\lbrk\bm c^{[j+1,k']}\rbrk_{k'\in \mathcal C_{j,k}} (\mathcal A^{[j,k]})^\top \\
&=\langle \lbrk\bm c^{[j+1,k']}\rbrk_{k'\in \mathcal C_{j,k}}, \mathcal A^{[j,k]}\rangle, \\
\end{aligned}
\]
and similarly, we have
\[
\begin{aligned}
\bm{d}^{[j,k]} & = \langle \bm  f, \bm \Psi^{[j,k]}\rangle 
=\lbrk\bm c^{[j+1,k']}\rbrk_{k'\in \mathcal C_{j,k}} (\mathcal B^{[j,k]})^\top \\
&=\langle \lbrk\bm c^{[j+1,k']}\rbrk_{k'\in \mathcal C_{j,k}}, \mathcal B^{[j,k]}\rangle.
\end{aligned}
\]
On the other hand, in view of the condition $(\mathcal A^{[j,k]})^\top \mathcal A^{[j,k]}+(\mathcal B^{[j,k]})^\top \mathcal B^{[j,k]} = \bm I$, together with \eqref{eq:dec1}, it is easy to see that \eqref{eq:rec1} holds.
\end{proof}

\begin{figure}[htpb!]
	\centering
	\subfloat[Forward]{\includegraphics[width=0.5\textwidth]{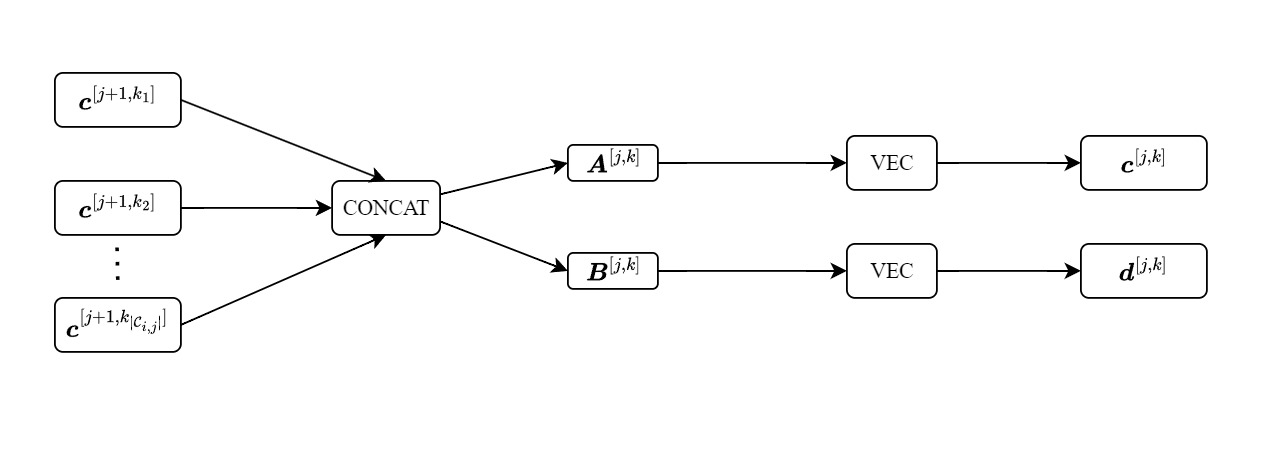}}
\subfloat[Backward]{\includegraphics[width=0.5\textwidth]{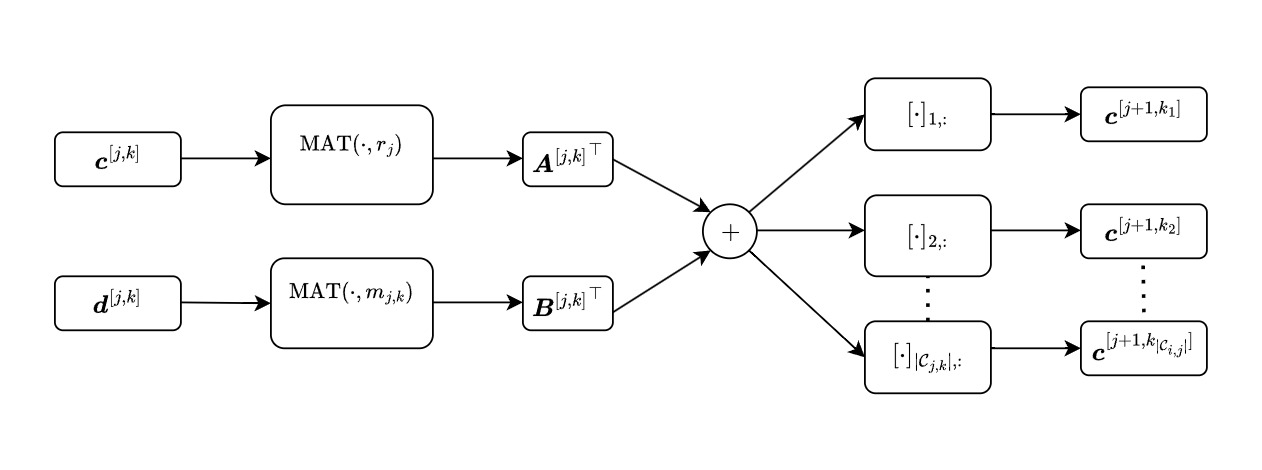}}
	\caption{(a) Forward transform: An illustration of computing coefficient vectors on a non-leaf node of $\mathscr{T}_J(\mathcal{G})$. The whole transform is computed iteratively in a bottom-up manner at each non-leaf node of $\mathscr{T}_J(\mathcal{G})$. (b) Backward Transform:An illustration of reconstruction on a non-leaf node of $\mathscr{T}_J(\mathcal{G})$. The whole transform is computed iteratively in a top-down manner at each non-leaf node of $\mathscr{T}_J(\mathcal{G})$. $[\cdot]_{i,:}$ denotes the operation of extracting the $i$-th row of the matrix. }
	\label{fig:filter}
\end{figure}

The one-level decomposition and reconstruction relations in Corollary \ref{cor_1} can be applied successively to decompose or reconstruct a graph signal $\bm f$. Such multi-level decomposition (forward transform)  and reconstruction (backward transform) procedures are summarized in Algorithms \ref{alg:Algo3} and \ref{alg:Algo4},  and illustrated in Figure~\ref{fig:filter}, respectively. Here, in Algorithm~\ref{alg:Algo4}, the notion $\text{ID}(k')$ is defined by sorting $k'\in \mathcal{C}_{j,k}$ increasingly to index each $k'\in \mathcal{C}_{j,k}$ by a unique integer $\text{ID}(k')$ in $\{1,\dots,|\mathcal{C}_{j,k}|\}$ such that $\text{ID}(k')< \text{ID}(k'') \Leftrightarrow k'\leq k''$.

\begin{algorithm}[htb!]\
		\caption{FORWARD $\mathsf{GIF}$-TRANSFORM}			           \label{alg:Algo3}
		\begin{algorithmic}[1]
			\State {\bfseries Input:} A signal $\bm{f}=[f_1,\ldots,f_n]\in L^2(\mathcal{G})$ and filter matrices $\bm{A}^{[j,k]}\in \mathbb{R}^{r_j\times |\mathcal{C}_{j,k}|},\bm{B}^{[j,k]}\in \mathbb{R}^{m_{j,k}\times |\mathcal{C}_{j,k}|},k\in[n_j],j\in\{0,\dots,J-1\}$ associated with a multi-graph tree partition $\mathscr{F}_J(\mathcal G)$. 
			\State $\bm{c}^{[J,k]}:={f}_k, k\in [n]$.
			\For{$j=J-1$ {\bfseries to} $0$}
			\For{$k=1$ {\bfseries to} $n_j$}
			\State $\bm{c}^{[j,k]}:=\text{\upshape{VEC}}(\bm{A}^{[j,k]}\text{\upshape{CONCAT}}(\{\bm{c}^{[j+1,k']}\}_{k'\in \mathcal{C}_{j,k}}))$
			\State $\bm{d}^{[j,k]}:=\text{\upshape{VEC}}(\bm{B}^{[j,k]}\text{\upshape{CONCAT}}(\{\bm{c}^{[j+1,k']}\}_{k'\in \mathcal{C}_{j,k}}))$
			\EndFor
			\EndFor
			\State {\bfseries Output:} $\{\bm{c}^{[0,1]}\}\cup\{\bm{d}^{[j,k]}\setsep k\in[n_j]\}_{j=0}^{J-1}$
		\end{algorithmic}
\end{algorithm}

\begin{algorithm}[htb!]\
		\caption{BACKWARD $\mathsf{GIF}$-TRANSFORM}			           \label{alg:Algo4}
		\begin{algorithmic}[1]
			\State {\bfseries Input:}  Coefficient vectors $\{\bm{c}^{[0,1]}\}\cup\{\bm{d}^{[j,k]}\setsep k\in[n_j]\}_{j=0}^{J-1}$ and  filter matrices $\bm{A}^{[j,k]}\in \mathbb{R}^{r_j\times |\mathcal{C}_{j,k}|},\bm{B}^{[j,k]}\in \mathbb{R}^{m_{j,k}\times |\mathcal{C}_{j,k}|},k\in[n_j],j\in\{0,\dots,J-1\}$.
			\For{$j=0$ {\bfseries to} $J-1$}
			\For{$k=1$ {\bfseries to} $n_j$}
			\State $\bm{X}:={\bm{A}^{[j,k]}}^\top\text{\upshape{MAT}}(\bm{c}^{[j,k]},r_j) +{\bm{B}^{[j,k]}}^\top\text{\upshape{MAT}}(\bm{d}^{[j,k]}\allowbreak, m_{j,k})$
			\For{$k'\in \mathcal{C}_{j,k}$}
			\State $\bm{c}^{[j+1,k']} := \bm{X}_{\text{\upshape{ID}}(k'),:}$
			\EndFor
			\EndFor
			\EndFor
			\State {\bfseries Output:} $\{\bm{c}^{[J,k]}\}_{k=1}^n$
		\end{algorithmic}
\end{algorithm}

\subsection{Computational Complexity}
We next discuss the computational complexity of Algorithms \ref{alg:Algo3} and \ref{alg:Algo4}. We have the following theorem.
\begin{theorem}
Adopt the notations in Theorem \ref{thm_1} and Corollary \ref{cor_1}. Assume that the computational complexity of $\text{CONCAT}$, $\text{VEC}$, and $\text{MAT}$ are of order $\mathcal{O}(\text{size of input})$, and there exist positive constants $t,s, r, \alpha, m$ such that $\min\{t, s, r, \alpha,m\}\ge 1$ and
\begin{gather}
t\leq |\mathcal{C}_{j,k}|\leq s,\, s\leq t^{\alpha}, \, r_j \leq r\le \frac{s}{2},\, m_{j,k}\leq m\label{com_eq_1}
\end{gather}
for all $k\in[n_j]$ and all $j\in\{0,\dots,J-1\}$.
Then the computational  complexity of Algorithm \ref{alg:Algo3} or \ref{alg:Algo4} is of order $\mathcal{O}((r+m) \cdot n^\alpha)$ 
\end{theorem}

\begin{proof}
By the tree structure of $\mathscr{T}_J(\mathcal V)$, we can deduce that
\begin{gather*}
J \leq \log_{t}(n),\, t^j\leq n_j\leq s^j, R_j\leq r^{J-j},\, j\in \{0,\dots,J\}.
\end{gather*}
For a fixed level $j$, the computational complexity at line 5 of Algorithm \ref{alg:Algo3} is of order
\begin{align*}
	&\mathcal{O}(s\cdot r^{J-j-1})&\quad&(\text{\upshape{CONCAT}}),\\
	+& \mathcal{O}(s\cdot r^{J-j})&\quad&(\text{\upshape{Matrix multiplication with }}\bm{A}^{[j,k]}),\\
	+ &\mathcal{O}(r^{J-j})&\quad&(\text{\upshape{VEC}}).
\end{align*}
Therefore, if excluding line 6, the total computational complexity is of order
\[
	\mathcal{O}\Big(\sum^{J-1}_{j=0}s^jr^{J-j}\left(sr^{-1}+s+1\right)\Big) = \mathcal{O}\Big(\sum^{J-1}_{j=0}s^jr^{J-j}S_1\Big)
\]
where $S_1:= sr^{-1}+s+1 = s(r^{-1}+1+s^{-1})\le 3s$. By (\ref{com_eq_1}) and the identity
$\sum_{j=0}^{J-1}x^{j} = \frac{x^J-1}{x-1}$, 
we have
$\sum_{j=0}^J s^jr^{J-j}=r^J\frac{(s/r)^J-1}{s/r-1} \le r^J\frac{(s/r)^J}{s/r-s/(2r)} \leq 2rs^{J-1}$
and the order of computational complexity becomes $\mathcal{O}(r\cdot s^{J-1}\cdot S_1)$, which is essentially of order $\mathcal{O}(r\cdot n^{\alpha})$ in view of $s\le t^\alpha$ and $J\le \log_{t}(n)$.

Similarly, at line 6 of Algorithm \ref{alg:Algo3}, if line 5 is excluded, the total computational  complexity is of order 
\[
\begin{aligned}
	&\mathcal{O}\Big(\sum^{J-1}_{j=0}s^jr^{J-j}\left(sr^{-1}+smr^{-1}+mr^{-1}\right)\Big)
=\mathcal{O}(rs^{J-1}\cdot S_2),
\end{aligned}
\]
where $S_2:=sr^{-1}+smr^{-1}+mr^{-1}\le 3smr^{-1}$, 
which is essentially of order $\mathcal{O}\left( m\cdot  n^{\alpha}\right)$

Combine the two terms we have order $\mathcal{O}\left((r+ m)\cdot n^{\alpha}\right)$ of computational complexity for Algorithm~\ref{alg:Algo3}.

As for Algorithm \ref{alg:Algo4}, it is obvious that lines 4, 5, and 6 together become the reverse of lines 5 and 6 of Algorithm \ref{alg:Algo3}, and that the amounts of computation are the same. Thus, the computational complexity of Algorithm \ref{alg:Algo4} is the same as that of Algorithm \ref{alg:Algo3}.
\end{proof}

\begin{remark}
{\rm
{The number $\alpha$ indicates the balance of children-node sizes $|\mathcal{C}_{j,k}|$.} If $\mathscr{T}_J(\mathcal{V})$ is a $K$-tree, i.e. each tree node has exactly $K$ children, then $\alpha = 1$ and we have order $\mathcal{O}((r+m)n)$ of computational complexity. 
{By definition, $m$ is an upper-bound of the numbers of rows of each $\bm{B}^{[j,k]}$. When the rows of $\bm{B}^{[j,k]}$ form a basis of $\Span{(\bm{B}^{[j,k]})}$, it is obvious that it is bounded above $|\mathcal{C}_{j,k}|$ and further bounded by $s$. Thus, $m$ can be $s$. When the rows of $\bm{B}^{[j,k]}$ form a frame of $\Span{(\bm{B}^{[j,k]})}$, we can control the number of elements of the frame to be a multiple of the dimension of $\Span{(\bm{B}^{[j,k]})}$ and thus $m$ can be a multiple of $s$. In both cases, we have $r+m = O(s)$ and the complexity further becomes $O(s\cdot n)$. In summary, constructing a partition tree with balanced and small children-node sizes ensures a low computation complexity. This is compatible with the complexity analysis in \cite{2023arXiv230604265L}.}
}
\end{remark}

\section{Data-adaptive Graph Framelet Systems}\label{sec:data-adaptive}

Given a multi-graph partition tree, we still need to specify the filter matrices $\bm{A}^{[j,k]},\bm{B}^{[j,k]}$ to obtain a graph-involved framelet system $\mathsf{GIF}_J(\mathcal{G})$.  In this section, we focus on learning such filter matrices from a certain family of signals through optimization techniques on Stiefel manifolds. For convienience, we call $\bm A^{[j,k]}$ and $\bm B^{[j,k]}$ associating with $\mathbb V_{j,k}, \mathbb W_{j,k}$ the \emph{low-pass filter} and \emph{high-pass filter}, respectively.

Suppose we are given $n_f$ sample graph signals $\bm{f}_1$, $\bm{f}_2$, $\dots$, $\bm{f}_{n_f}$ of a family $\mathfrak{F}$ with low-dimensional structure, i.e., the family $\mathfrak{F}$ is approximately contained in a low-dimensional linear space in $\mathbb{R}^n$. Similar to the setting in machine learning tasks, we specify the parameters (filters) by learning from these samples and expect that the learned $\mathsf{GIF}_J(\mathcal{G})$ will be able to provide sparse representation for any signal in $\mathfrak{F}$.

For determining the low-pass filters, we aim at learning low-pass filters so that the space $\mathbb V_0=\Span(\bm\Phi_0)$ provides the best fitting of the samples in the $l^2$ sense. That is, for any signal $\bm{f} \in \mathfrak{F}$, most of the energy of $\bm{f}$ is contained in its projection on $\mathbb V_0$. In such a way, we could capture the low-dimensional structure of the family $\mathfrak{F}$. Once we obtained the low-pass filters, in view of the generalized vanishing moments, i.e., orthogonality between $\mathbb V_0$ and $\mathbb W_j$, most of the coefficients in $\mathbb W_j$ are expected to be small or even zero. Therefore, in such a case, the $\mathcal{G}$-framelet system  $\mathsf{GIF}_J(\mathcal{G})$ will provide sparse representations for all signals in $\mathfrak{F}$ (after thresholding).

The collection of matrices in $\mathbb{R}^{n\times p }$ such that $\bm{M}^\top\bm{M}=\bm{I}$ forms a \emph{Stiefel manifold} \cite{absil2008optimization} endowed with the Riemannian metric induced from the Euclidean metric of $\mathbb{R}^{n\times p}$. We denote such a manifold in $\mathbb{R}^{n\times p}$ by $\mathcal{M}^{[n,p]}$. Next, we provide the learning of the low- and high-pass filters through optimization on Stiefel manifolds.

\subsection{Learning the Low-pass Filters}\label{subsec_lowpass}
We can regard the filters $\bm{A}^{[j,k]},\bm{B}^{[j,k]}$ as \emph{parameters} of $\mathsf{GIF}_J(\mathcal{G})$, and it can be determined by certain criteria. These filters are similar to the weights in a neural network, except that each $\bm{A}^{[j,k]}$ are points on a Stiefel manifold $\mathcal{M}^{[r_j,|\mathcal{C}_{j,k}|]}$ and that the rows of $\bm{B}^{[j,k]}$ are vectors in the orthogonal complement of the space spanned by the rows of  $\bm{A}^{[j,k]}$. 

More specifically,  assuming $r_j, j=0,\ldots, J-1$ are specified such that  $r_j \in [\min(|\mathcal{C}_{j,1}|,\dots,|\mathcal{C}_{j,n_j})|-1]$, we solve the following problem for the low-pass filters:
\begin{equation}\label{opt_1}
\underset{{\left\{\bm{A}^{[j,k]}\in \mathcal{M}^{[r_j,|\mathcal{C}_{j,k}|]}\setsep k\in[n_j]\right\}_{j=0}^{J-1}} }{\arg\max}
\Vert \bm{c}^{[0,1]}_1 \Vert^2  + \dots + \Vert \bm{c}^{[0,1]}_{n_f} \Vert^2,\\
\end{equation}
where $\bm{c}^{[0,1]}_i=\langle \bm f_i, \bm\Phi_0\rangle$, $i\in[n_f]$, are the coefficients with respect to $\bm{\Phi}_0$ for $\bm{f}_i$, $i\in[n_f]$, computed as in Algorithm \ref{alg:Algo3}. Note that these coefficients only depend on the filters $\bm{A}^{[j,k]}$. By solving the problem in \eqref{opt_1}, the scaling functions $\bm{\Phi}_0$ and the space $\mathbb V_0$ are simultaneously determined. Moreover, the objective function in \eqref{opt_1} is smooth on the product manifold $\mathcal{M}_{\bm A}:=\otimes_{j,k}\mathcal{M}^{[r_j,|\mathcal{C}_{j,k}|]}$ formed by all $\mathcal{M}^{[r_j,|\mathcal{C}_{j,k}|]}$. Thus, the problem above can be regarded as an optimization on Riemannian manifolds and we can choose any Riemannian optimization algorithm that is eligible to apply, e.g., see \cite{absil2008optimization}. 

By definition \eqref{opt_1} gives the best orthonormal vectors that best fit the training signals in $l^2$ sense under the constraint that they are induced by the filters $\bm{A}^{[j,k]}$. Let $\bm{M} := \llbracket \bm{f}_i\rrbracket_{i\in [n_f]}$ denote the matrix formed by concatenating the training signals. Without requiring the orthonormal vectors are the rows of $\bm \Phi_0$, (\ref{opt_1}) becomes 
\[
    \underset{\bm{P} \in \mathbb{R}^{R_0\times n}, \bm{P}\bm{P}^\top = \bm{I}}{\arg\max} \Tr(\bm{P}\bm{M}^{\top}\bm{M}\bm{P}^{\top}).
\]
This is a well-known problem which is related to principal component analysis, and one of the optimal solutions is the first $R_0$ right singular vectors of $\bm{M}$ (see e.g. the Eckart-Young theorem \cite{golub2013matrix}). Hence, other than conducting Riemannian optimization for (\ref{opt_1}), another possible way to decide the filters $\bm{A}^{[j,k]}$ is to obtain the first $R_0$ right singular vectors of $\bm{M}$, then decide the entries in $\bm{A}^{[j,k]}$ such that $\Phi_0$ contains exactly the first $R_0$ right singular vectors of $\bm{M}$. However, this is guaranteed for some special cases, which needs the following lemma.
\begin{lemma}\label{new_prop}
Let $\text{supp}(\bm p)$ denote the support of a vector $\bm{p}$ in $\mathbb{R}^n$, i.e. the indices with non-zero values and $\bm{p}|_{I},I\subseteq [n]$ denote the restriction of $\bm{p}$ on a set $I$ of indices. Given vectors $\bm{p},\bm{q_1},\dots,\bm{q}_r \in \mathbb{R}^n$ with unit $l^2$ norm, if $\bigcup_{i=1}^m \text{supp}(\bm{q}_i) = \text{supp}(\bm{p})$ and $\text{supp}(\bm{q}_i)\cap\text{supp}(\bm{q}_j) = \varnothing, i\neq j$. Then $\bm{p} = \sum_{i=1}^m a_i\bm{q}_i$ if and only if
\[
    \bm{q}_i = \frac{\bm{p}|_{\text{supp}(\bm{q}_i)}}{\Vert \bm{p}|_{\text{supp}(\bm{q}_i)}\Vert},\quad 
    a_i = \Vert \bm{p}|_{\text{supp}(\bm{q}_i)}\Vert,\quad i= 1,\dots,m.
\]
\end{lemma}
\begin{proof}
The \textit{if} part is straightforward. The \textit{only if} part can be implied from the conditions $\Vert\bm{q}_i\Vert = 1$ and $\bigcup_{i=1}^m \text{supp}(\bm{q}_i) = \text{supp}(\bm{p})$ and $\text{supp}(\bm{q}_i)\cap\text{supp}(\bm{q}_j) = \varnothing, i\neq j$.  
\end{proof}
Then, we have the following theorem. 
\begin{theorem}
Let $\mathscr{T}_J({\mathcal{V}})$ be a partition tree as in Definition~\ref{def1}. Fix $r_j =1$ for $j=0,\dots,J-1$ in Definition~\ref{def2}. Let $\bm{p}$ be an orthonormal vector. Then, there exist filter matrices $\bm{A}^{[j,k]}\in \mathbb{R}^{r_j\times |\mathcal{C}_{j,k}|}$ associated with the tree node $p_{j,k}$ for $j=0,\ldots, J-1$ and $k\in[n_j]$, such that $\bm\Phi_0$ in the induced $\mathcal V$-framelet system $\mathcal{F}_{0}^J(\mathcal V)$ contains exactly $\bm{p}$.
\end{theorem}
\begin{proof}
Since $r_j = 1$, the subspaces $\mathbb{V}^{[j,k]}$ are one-dimensional, and each $\bm \Phi^{[j,k]}$ can contain only one row. Note that in this case, $\mathcal A^{[j,k]} = \bm{A}^{[j,k]}$ and $ \bm{A}^{[j,k]}$ is a row vector. Starting from $\bm \Phi^{[0,1]} = [\bm{p}]$, if for any $\mathcal{S}_{1,k}$, $\mathcal{S}_{1,k}\cap\mathrm{supp}(\bm{p}) = \varnothing$, then let $\bm{\Phi}^{[1,k]}$ be any unit-norm row vector with support contained in $\mathcal{S}_{1,k}$, and the corresponding entry in $\bm{A}^{[j,k]}$ be zero; for the remaining $\mathcal{S}_{1,k}$ such that $\mathcal{S}_{1,k}\cap\mathrm{supp}(\bm{p}) \neq \varnothing$, by (\ref{AB:jk}) and Lemma \ref{new_prop}, we can decide the entries of $\bm{A}^{[0,0]}$ and the matrices $\bm \Phi^{[1,k]}$, which correspond to the $a_i$ and $\bm{q}_i$, respectively. We can recursively perform such deductions from top to bottom until all $\bm{A}^{[j,k]}$ and $\bm{\Phi}^{[j,k]}$ are decided.
\end{proof}
 Therefore, when all the subspaces $\mathbb{V}_{j,k} $ are one-dimensional, i.e., we only seek one generalized vanishing moment, we can decide all the filters $\bm{A}^{[j,k]}$ in a top-down manner. However, if some of the subspaces $\mathbb{V}_{j,k}$ are not one-dimensional, then such a procedure fails. For example, suppose that the dimensions of $\mathbb{V}_0$ and all subspaces $\mathbb{V}_{1,k}$ are $2$ and that the basis of $\mathbb{V}_0$ is known. In this case, the filter $\bm{A}^{[0,0]}$ is essentially a row vector. By taking the first vector in the basis of $\mathbb{V}_0$ and applying Proposition \ref{new_prop}, the filter $\bm{A}^{[0,0]}$ and the first vectors in the bases of each $\mathbb{V}_{1,k}$ are decided. However, doing the same for the second vector in the basis of $\mathbb{V}_0$ produces a (generally) different set of values for $\bm{A}^{[0,0]}$, which leads to a contradiction. When only $V_0$ is multi-dimensional, the conclusion is similar, in which Lemma \ref{new_prop} generally leads to different sets of $\bm\Phi^{[1,k]}$. However, we can have one fixed set. In such cases, we resort to solving \eqref{opt_1}. Such differences in learning the filters $\bm{A}^{[j,k]}$ actually present a trade-off in our systems: allowing the subspaces $\mathbb{V}_{j,k}$ to have more than one dimension gives the possibility to obtain multi-dimension $\mathbb{V}_0$ using fewer parameters in $\bm{A}^{[j,k]}$, with the price of losing algorithmic simplicity and optimality.

\subsection{Learning the High-pass Filters}
Once the filters $\bm{A}^{[j,k]}$ are determined, in view of the condition in  \eqref{eq3:thm1c}, we can obtain each $\bm{B}^{[j,k]}$ by finding a basis or a frame of the orthogonal complement of the rows of $\bm{A}^{[j,k]}$. For the former case, we can apply singular value decomposition (SVD). For the latter case, we can apply the method of constructing undecimated framelet systems in \cite[Theorem 2]{zheng2022decimated} on a set of orthonormal eigenvectors and select three filters $\bm{\alpha},\bm{\beta}_1,\bm{\beta}_2$ as in \cite[Section 4.1]{wang2020tight}. Here, we choose not to describe the details for brevity and assume that such \emph{initial high-pass filters} $\bm{B}^{[j,k]}$ are obtained either through the SVD technique or the undecimated framelet approach. We would like to emphasize that any method for constructing frames on the span of a set of orthonormal vectors is eligible.

It is obvious that the high-pass filters $\bm{B}^{[j,k]}$ are not unique. By multiplying $\bm{B}^{[j,k]}$ with an orthogonal matrix $\bm{U}^{[j,k]} \in \mathbb{R}^{m_{j,k}\times m_{j,k}}$ on the left, we obtain a new $\mathcal{G}$-framelet system $\mathsf{GIF}_J(\mathcal{G})$ with the same subspaces $\mathbb W_{j,k}$ and preserving the  tightness property. Thus, given $\bm{A}^{[j,k]},\bm{B}^{[j,k]}$, to promote the sparsity of $\mathsf{GIF}_J(\mathcal{G})$, for each $\mathbb W_{j,k}$, we consider solving the following problem:
 \begin{equation}\label{opt_2}
\begin{small}
\begin{gathered}
\underset{\bm{U}^{[j,k]} \in \mathcal{M}^{[m_{j,k},m_{j,k}]}}{\arg\min}\ \Vert \bm{d}^{[j,k]}_1(\bm{U}^{[j,k]})^\top \Vert_1 
+ \dots + \Vert \bm{d}^{[j,k]}_{n_f}(\bm{U}^{[j,k]})^\top \Vert_1,\\
\end{gathered}
\end{small}
\end{equation}
where $\bm{d}^{[0,1]}_i=\langle \bm f_i, \bm\Psi^{[j,k]}\rangle$, $i\in[n_f]$ are the coefficients with respect to $\bm{\Psi}^{[j,k]}$ for $\bm{f}_i$, $i\in[n_f]$, computed as in Algorithm \ref{alg:Algo3} and $\Vert \cdot \Vert_1$ is the $l_1$ norm. Note that $\bm{d}^{[j,k]}_1(\bm{U}^{[j,k]})^\top$ are the coefficients in $\mathbb W_{j,k}$ with $\bm{B}^{[j,k]}$ replaced by $\bm{U}^{[j,k]}\bm{B}^{[j,k]}$. Therefore, the problem in \eqref{opt_2} aims at learning the sparsest basis/frame for a fixed $\mathbb W_{j,k}$. To solve \eqref{opt_2},  we follow \cite{trendafilov2003} and substitute the $l_1$ norm $\Vert \bm{M}\Vert_1$ with a smooth surrogate
\[
	\Tr\left(\bm{M}^\top \tanh(1000\cdot\bm{M})\right),
\]
where $\tanh$ is applied element-wise. In such a way, similar to  \eqref{opt_1},  the problem in \eqref{opt_2} becomes a smooth objective on the Stiefel manifold. Moreover,  to reduce the optimization overhead, we do not solve \eqref{opt_2} for all $k\in[n_j]$ and all $j=0,\ldots, J-1$. In fact, according to the total $l^2$ norm of the coefficients, we only solve \eqref{opt_2} for certain $j,k$ by choosing among those $\mathbb W^{[j,k]}$ the top-$N$ subspaces to optimize. More specifically, we sort the subspaces $\mathbb{W}_{j,k}$ decreasingly according to the initial coefficient quantities $\epsilon^{[j,k]}$:
\[
	\epsilon^{[j,k]}:=\Vert\bm  d_1^{[j,k]}\Vert^2 + \Vert \bm d_2^{[j,k]}\Vert^2 + \dots + \Vert \bm d_{n_f}^{[j,k]}\Vert^2,
\]
and solve the problem \eqref{opt_2}
only for those $j,k$  with respect to the first $N$ largest $\epsilon^{[j,k]}$.

To facilitate the descriptions in the next section, we use $M$ to denote the dimension of $\mathbb V_0$ and $\mathsf{GIF}$-I$(M)$ to denote a $\mathcal{G}$-framelet system obtained by solving only problem \eqref{opt_1} and constructing  $\bm{B}^{[j,k]}$ through the undemicated framelet system in \cite[Section 4.1]{wang2020tight}. Similarly, we use $\mathsf{GIF}$-II$(M,N)$ to denote a $\mathcal{G}$-framelet system obtained by solving both problem \eqref{opt_1} and the modified version of problem \eqref{opt_2} on the top-$N$ spaces with respect to the first $N$ largest $\epsilon^{[j,k]}$. When all $\bm{B}^{[j,k]}$ are constructed as bases (from SVD), we denote them slightly differently as $\mathsf{GIB}$-I$(M)$ and $\mathsf{GIB}$-II$(M, N)$, respectively.

We conclude this section by some further remarks on the differences between our method and two closely related works \cite{sharon2015class, nakahira2018parseval}.
\begin{remark}
{\rm If we replace the first $k$ Laplacian eigenvectors and apply the same procedure in \cite{sharon2015class}, the output will be very similar to ours despite the former being restricted to a basis. Thus, our system seems to be just replacing $k$ Laplacian eigenvectors with $k$ arbitrary orthonormal vectors. In fact, there are some important differences. In \cite{sharon2015class}, the wavelets are constructed on the partition tree in \textbf{top-down} manner, in which the number $k$ affects the shape of the partition tree. In contrast, our method starts from a given partition tree and assigns parameters to each node in the partition in a \textbf{bottom-up} manner. We can adjust the size of parameters so that the number of scaling functions $k$ and the shape of the partition tree are decoupled to a certain extent. This is important since the shape of the partition tree determines the sparseness of the framelets, and we do not want the number $k$ to affect the sparseness. Once the sizes of the parameters are fixed, as shown in Section \ref{subsec_lowpass}, we can determine the values of the parameters in a top-down manner similar to \cite{sharon2015class} in some special cases.
}
\end{remark}

\begin{remark}
{\rm 
\cite{nakahira2018parseval} proposes a framework to construct Parseval frames on directed acyclic graphs. Since the partition tree is an instance of directed acyclic graphs (DAGs), we can essentially produce the same framelets and framelet transforms using the systems in \cite{nakahira2018parseval} (with slight modifications). However, the notation and data structures adopted in \cite{nakahira2018parseval} are relatively complicated. In detail, a DAG must be converted into a direct acyclic graph with joint points (DAGJP). A wavelet transform on hierarchical graphs (WTHG), which is a computational diagram depicting the procedure of computing wavelet coefficients, is further obtained by converting the DAGJP. In short, it does not provide comparably simple and concise notations as in classical 1-D framelets (see \cite[Chapter 1]{han2017framelets}) due to its generality. In contrast, when confined to partition trees, our work provides notations and transforms that are more concise and simpler without losing much of the generality.
On the other hand, optimizations for unitary transform matrices in \cite{nakahira2018parseval} are conducted layer-by-layer to obtain sparseness. This is different from the procedure above.}
\end{remark}


\section{Graph Signal Processing}\label{sec:expr}
In this section, we demonstrate the efficiency and effectiveness of our graph framelet systems on graph signal processing through several numerical experiments. 

\subsection{Data Preparation and Implementation Details}
Here we consider defining a synthetic family of signals on the common benchmark dataset \textit{Minnesota traffic graph} \cite{Hammond2011wavelets}. To facilitate comparisons with GraphSS \cite{sakiyama2019two}, where the number of nodes is a multiple of 2 to enable decomposition, we follow the publicly released code of GraphBior \cite{narang2013compact} in which the connected component formed by node number 347 and 348 is erased, resulting in a connected graph of $2640$ nodes. The family $\mathfrak{F}$ of signals is defined as follows: 
\begin{enumerate}

\item{Find a minimal spanning tree of the Minnesota traffic graph. Pick a node as the root.}

\item{Generate a sum of $10$ Chebyshev polynomials $T(x):=\sum_{i=0}^9 a_iT_i(x)$ \cite{freud2014orthogonal}, where each $a_i$ is sampled from Gaussian distribution.}

\item{Generate $n_{d}$ equally-spaced points on interval $[-1,1]$, where $n_{d}$ is the depth of of spanning tree.}

\item{Assign each node with value $T(x_{v_d})$ where $v_d$ is the depth of the node in the spanning tree and $x_{v_d}$ is the $(v_d+1)$-th equally-spaced point on $[-1,1]$.}

\end{enumerate}

\begin{figure*}
\centering
%

\subfloat[Minnesota traffic]{\includegraphics[width=0.25\textwidth]{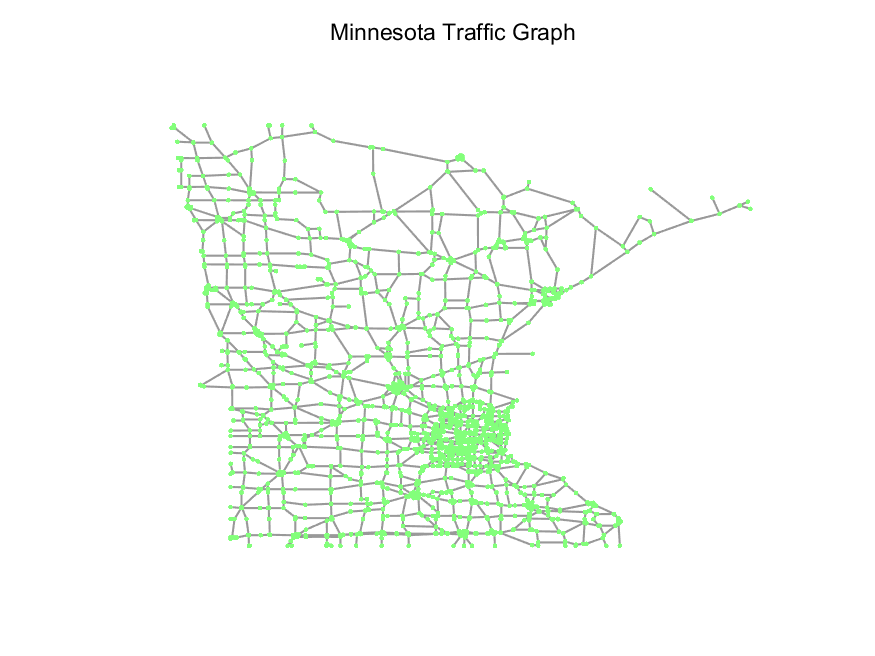}}
\subfloat[Spanning tree]{\includegraphics[width=0.25\textwidth]{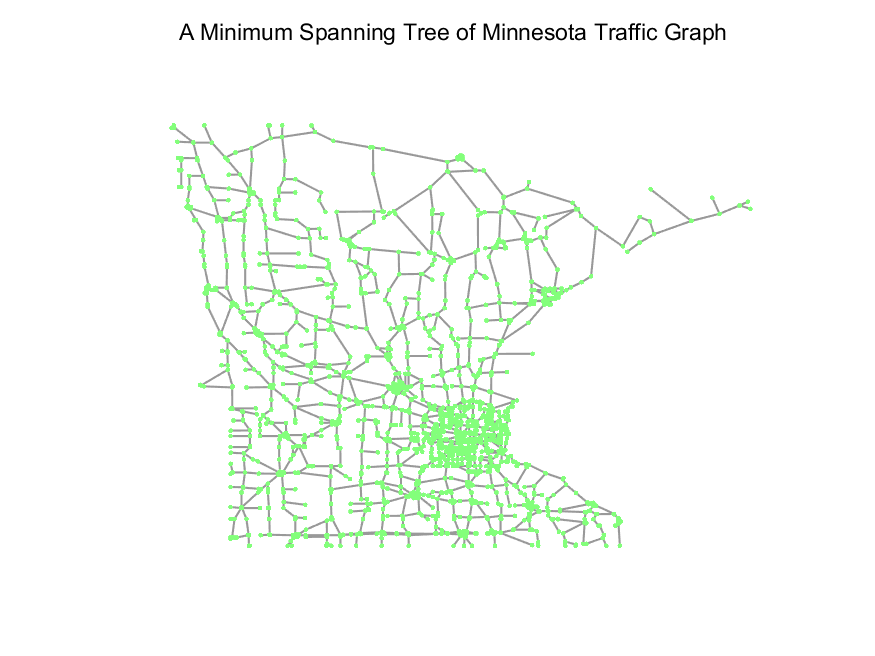}}
 \subfloat[Sample 1]{\includegraphics[width=0.25\textwidth]{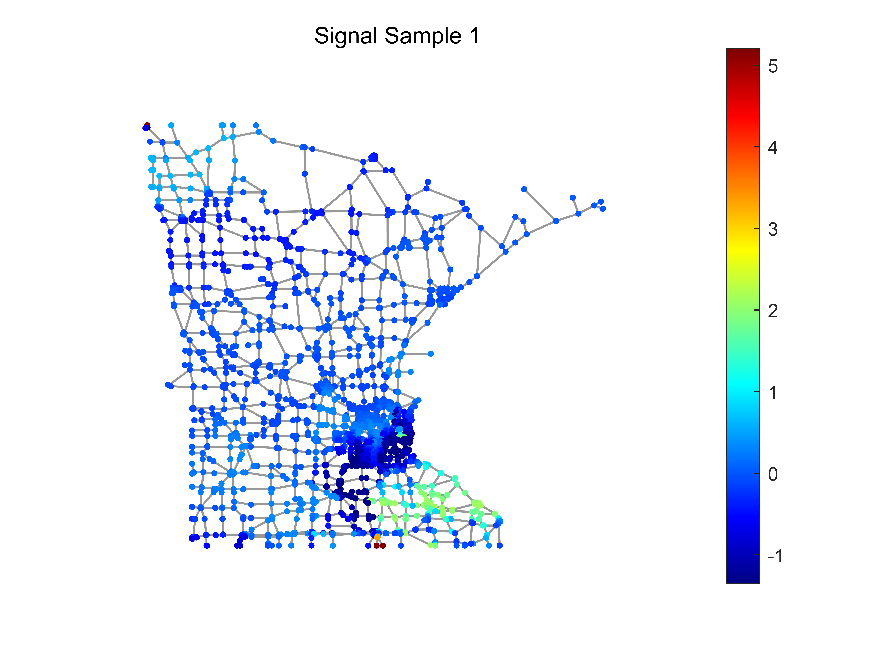}}
 \subfloat[Sample 2]{\includegraphics[width=0.25\textwidth]{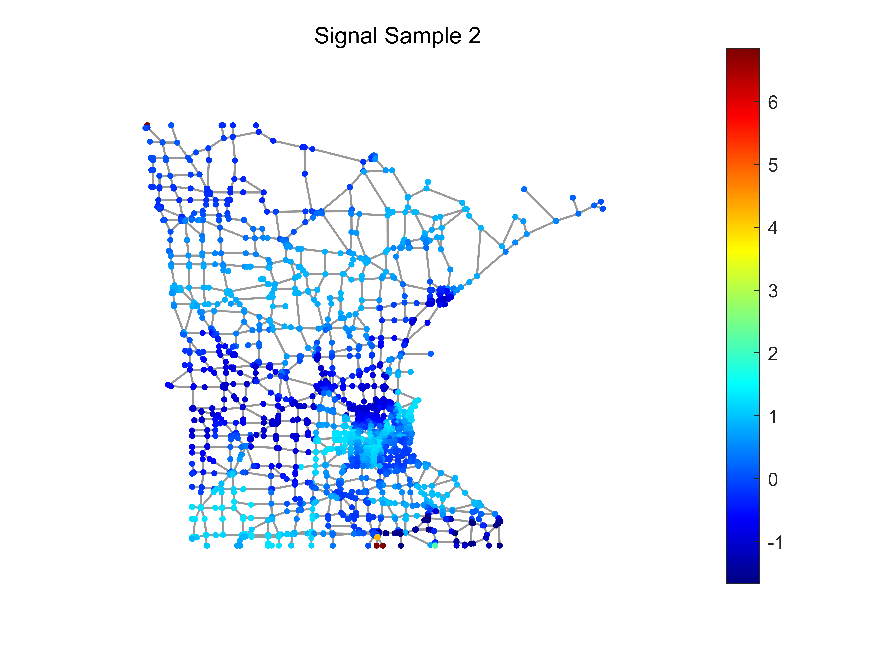}}
\caption{An illustration of (a) the Minnesota traffic graph, (b) one of its minimum spanning trees, and (c-d) 2 signal samples from the family $\mathfrak{F}$.}
\label{fig:signal}
\end{figure*}

Signals from this family are generated by sampling $a_i$ from the Gaussian distribution and following the evaluation as above (see Figure \ref{fig:signal}). We generated $50$ signals for training and $5$ signals for testing.

For the following tasks of non-linear approximation and denoising, we choose the following methods as baselines: 
\begin{itemize}
\item[1)] Eigenvectors of unnormalized graph Laplacian $\bm{L}:=\bm D-\bm W$ (abbreviated as UL). 

\item[2)] Eigenvectors of normalized graph Laplacian $\tilde{\bm{L}}:= \bm{I} - \bm{D}^{-1/2}\bm{W}\bm{D}^{-1/2}$  (abbreviated as NL). 

\item[3)] GraphSS \cite{sakiyama2019two}. 

\item[4)] GraphBior \cite{narang2013compact}. 

\end{itemize}
For GraphSS and GraphBior, we adopted the publicly released MATLAB code. In detail, we selected ``orth'' filters and a 3-level decomposition of GraphSS, which resulted in an orthonormal basis. For GraphBior, we used the same configuration as in the file \textit{Biorth}\_\textit{filter bank}\_\textit{demo}\_2.\textit{m}, which resulted in a biorthogonal basis.

Our method was implemented with Python\footnote[1]{https://github.com/zrgcityu/Graph-Involved-Frame}. To construct a partition tree for the Minnesota traffic graph, we adopted the Python package \textit{scikit-network}\footnote[2]{https://scikit-network.readthedocs.io} and applied a step of post-processing to perform a series of clustering on graphs as described at the beginning of Section \ref{sec:gif}. Eventually, we obtained a partition tree with $J=3$, in which each non-leaf node had no less than $15$ children, i.e. $|\mathcal{C}_{j,k}|\geq 15$. This allows choosing large enough $r_j$ for $\bm{A}^{[j,k]}$ and thus allows varying the dimension of $\mathbb V_0$. To obtain a $M$-dimensional $\mathbb V_0$, we set $r_{J-1} = M$ and the rest $r_j$ were set to $1$. As for solving the problem in \eqref{opt_1} and the modified version of the problem in \eqref{opt_2}, we adopted the Python package \textit{pymanopt}\footnote[3]{https://pymanopt.org/}\cite{townsend2016pymanopt} and applied the conjugate gradient method on Riemannian manifolds. The configuration was set to default, and a summary of the time consumed is presented in Table \ref{tab:1}. Figures of graphs were generated using GSPBOX\footnote{https://epfl-lts2.github.io/gspbox-html/} \cite{perraudin2014gspbox}.

\begin{table}[htpb!]
	\centering
	\scriptsize
	\caption{Time consumed on Riemannian optimizations. Unit: second.}
	\label{tab:1}
	\begin{tabular}{rl|rl|rl}
		\toprule
		Variant & Time  & Variant & Time & Variant & Time \\
        \midrule
		$\mathsf{GIB}$-I$(1)$  & 1002.03 & $\mathsf{GIB}$-II$(1,20)$ & 1158.92 & $\mathsf{GIB}$-II$(1,100)$ & 1772.85\\
		$\mathsf{GIB}$-I$(4)$  & 1001.59 & $\mathsf{GIB}$-II$(4,20)$ & 1147.13 & $\mathsf{GIB}$-II$(4,100)$ & 1629.42\\
		$\mathsf{GIB}$-I$(8)$  & 1002.45 & $\mathsf{GIB}$-II$(1,50)$ & 1388.68 & $\mathsf{GIF}$-II$(1,20)$ & 1366.11\\
		$\mathsf{GIB}$-I$(12)$ & 1002.44 & $\mathsf{GIB}$-II$(4,50)$ & 1350.72 & $\mathsf{GIF}$-II$(4,20)$ & 1307.43\\
		\bottomrule
	\end{tabular}
\end{table}

\subsection{Non-linear Approximation}
In this subsection, we choose only bases constructed from our method, e.g., $\mathsf{GIB}$-I$(4)$. To conduct non-linear approximation (NLA) on a signal $\bm{f}$ using top-$N$ coefficients of a dictionary $\mathcal{D} := \{\bm{g}_1,\bm{g}_2,\dots,\bm{g}_n\}\subset \mathbb{R}^n$, we sorted the coefficients $c_i := \langle \bm{f},\bm{g}_i \rangle$ decreasingly according to $|c_i|$ and kept the top-$N$ coefficients. The remaining ones were set to $0$. Then a signal $\tilde{\bm{f}}$ was reconstructed according to these modified coefficients using the proper inverse operator. We report the relative error, i.e., $\Vert \bm{f} - \tilde{\bm{f}}\Vert/\Vert \bm{f}\Vert$.

\subsubsection{Comparison among Variants}
\begin{figure}[htp]
\centering
\subfloat[]{\includegraphics[width=0.33\textwidth]{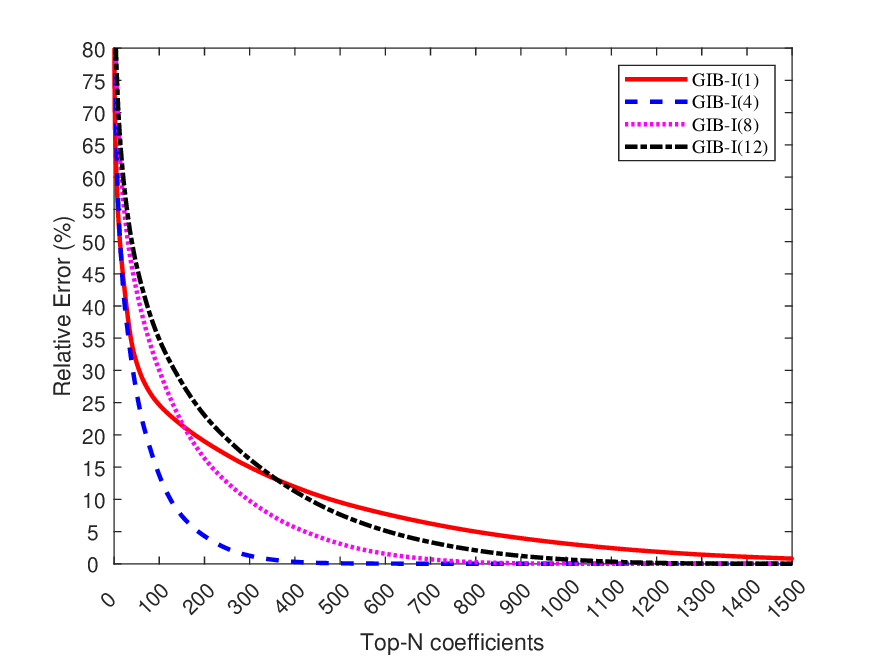}}
\subfloat[]{\includegraphics[width=0.33\textwidth]{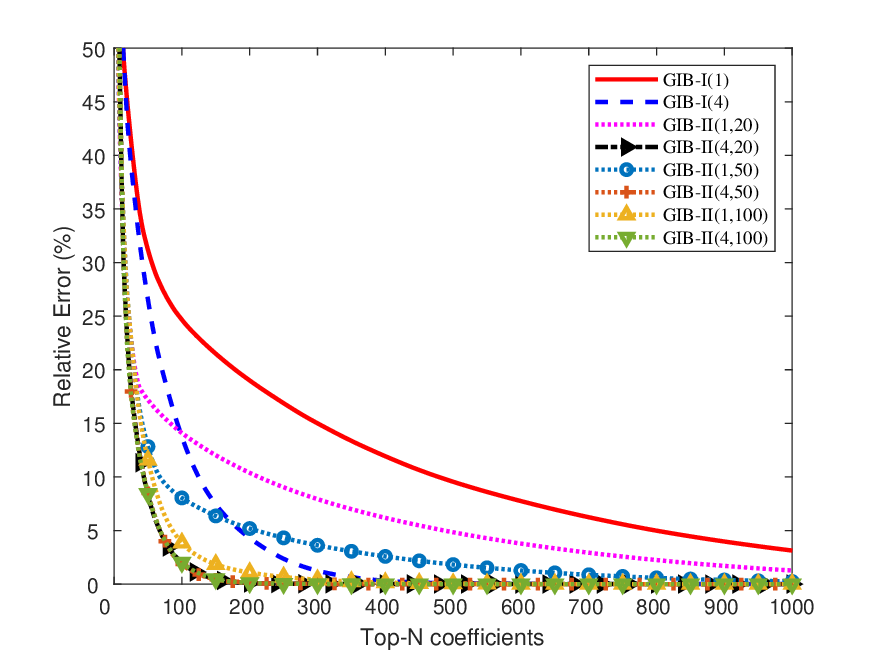}}
\subfloat[]{\includegraphics[width=0.33\textwidth]{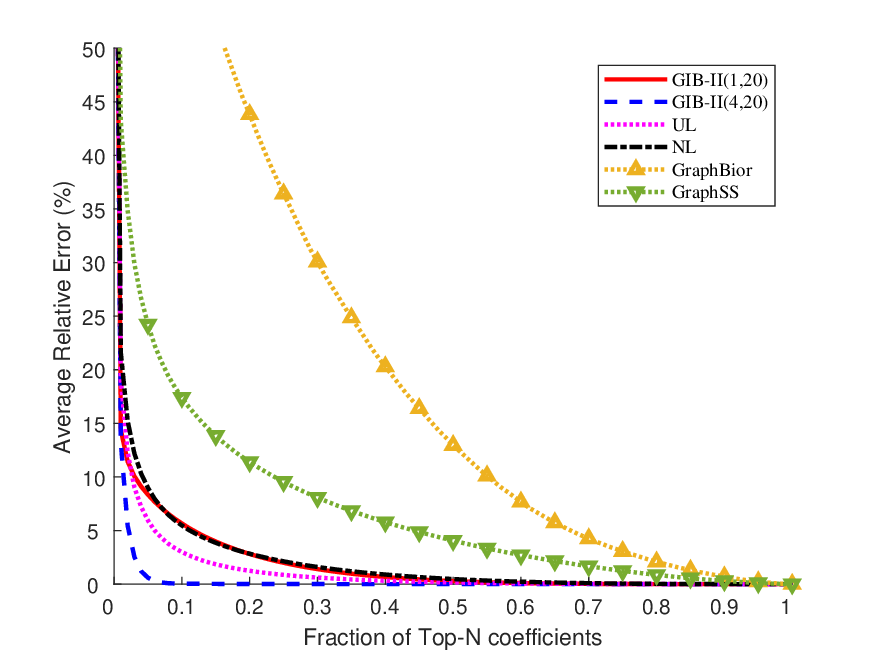}}
\caption{(a) Relative error of 4 variants of $\mathsf{GIB}$-I. (b) Relative error of 8 variants of $\mathsf{GIB}$-I and $\mathsf{GIB}$-II.
 (c)  Average relative error of all methods.}
\label{fig:com_1}
\end{figure}

Here, we compare the results among variants with different dimensions of $\mathbb V_0$. The NLA is performed on one of 5 test signals. As shown in Figure \ref{fig:com_1} (a), when the dimension is $4$, the learned $\mathbb V_0$ best fit the family $\mathfrak{F}$ and produces the sparsest representation for the test signal, indicating a low-dimensional structure of $\mathfrak{F}$ with a dimension larger than 1. On the other hand, for the cases of dimensions 8 and 12, the results indicate a surplus in defining the size of $\mathbb V_0$, which distributes the energy of the signal over an excessive set and thus results in a denser representation. In contrast, the one-dimensional case generally has the densest representation since a one-dimensional $\mathbb V_0$ is insufficient in fitting the low-dimensional structure of $\mathfrak{F}$.

Given the observation above, we further compare the two extreme cases, i.e. $\mathsf{GIB}$-I$(1)$ and $\mathsf{GIB}$-I$(4)$, for which optimize the filters $\bm{B}^{[j,k]}$ as well. The NLA is performed on the same test signal. As shown in Figure \ref{fig:com_1} (b), optimizing $\bm{B}^{[j,k]}$ gives sparser representations. For $\mathsf{GIB}$-I$(4)$, the results of optimizing top-20, 50, and 100 subspaces are indistinguishable since the four-dimensional $\mathbb V_0$ has conserved most of the energy and thus it requires only a small portion of subspaces to optimize. 

On the contrary, for $\mathsf{GIB}$-I$(1)$, there is a great portion of energy distributed outside of the one-dimensional $\mathbb V_0$. Thus, it requires more subspaces to optimize. Only when 100 subspaces are optimized, i.e., $\mathsf{GIB}$-I$(1,100)$, does the one-dimensional case achieve a performance comparable with the four-dimensional case, i.e., $\mathsf{GIB}$-II$(4,20)$. 

In conclusion, selecting a proper order of generalized vanishing moments at the beginning provides more efficient and economic optimization in obtaining $\bm{B}^{[j,k]}$.


\subsubsection{Comparison with Baselines}
Based on the aforementioned results, we select $\mathsf{GIB}$-II$(1,20)$ and $\mathsf{GIB}$-II$(4,20)$ to compare with the baselines. We report the average relative error of the 5 test signals. As shown in Figure \ref{fig:com_1} (c), $\mathsf{GIB}$-II$(4,20)$ evidently outperforms all other methods.


\subsection{Denoising}
In this subsection, we include two more variants $\mathsf{GIF}$-II$(1,20)$ and $\mathsf{GIF}$-II$(4,20)$ to compare with the baselines.

For each test signal, we added Gaussian noise with $\sigma = 1/16,1/8,1/4$ and $1/2$. Then, the coefficients were thresholded with a threshold value set as  $3\sigma$. We report the signal-to-noise ratio (SNR). The results are shown in Table \ref{tab:2}. 

Our method surpasses the baselines by a large margin. A sparser representation is better for denoising since more coefficients are dominated by noise, and thus thresholding such coefficients results in less loss of the original signal. Moreover, we mentioned in the introduction that frames have the potential to perform better since the energy of noise is distributed over more atoms, resulting in smaller coefficients. With the current setting, as shown in the last column of Table \ref{tab:2}, this happens when the noise is large enough. In other cases, their performance is very close to the bases.

\begin{table}[htpb!]
	\centering
	\scriptsize
	
	\caption{Average SNR. Best in \textbf{bold}. Unit: dB.}
	\label{tab:2}
	\begin{tabular}{lllll}
		\toprule
		Method & $\sigma = 1/16$ & $\sigma = 1/8$ & $\sigma = 1/4$ & $\sigma = 1/2$ \\ \midrule
		UL	&17.97	&13.92	&10.46	&7.54\\
		NL	&16.51	&12.29	&9.2	&6.54\\
		GraphBior	&9.59 &6.06& 2.76	&0.88\\
		GraphSS	&8.92	 &6.18	&3.73  &1.76\\
		$\mathsf{GIB}$-II$(1,20)$	&17.32	&12.37	&9.49	 &7.58\\
		$\mathsf{GIB}$-II$(4,20)$	& \textbf{25.31}	& \textbf{20.15}	& \textbf{14.67}	&9.62\\
		$\mathsf{GIF}$-II$(1,20)$	&16.6 	&12.16	&9.6	&7.91\\
		$\mathsf{GIF}$-II$(4,20)$	&25.24	&20.03	&14.67	& \textbf{10.32}\\

		\bottomrule
	\end{tabular}
\end{table}

\section{Node Classification}\label{sec:gnn}

\subsection{Graph Framelets as Features}\label{gnn_subsec_1}

\par As mentioned in Section \ref{intro_subsec_1}, we are interested in generating graph framelets to encode structural information other than the adjacency matrices. As the adjacency matrices are simply the connections between nodes and the neighbors that are $1$-hop away, we instead look at structural information that is beyond $1$-hop. To do so, we first introduce the induced two-hop graph. In detail, 
an induced two-hop graph $\mathcal{G}_{\text{2hop}}:=(\mathcal{V}^{\text{2hop}},\mathcal{E}^{\text{2hop}},\bm{W}^{\text{2hop}})$ of a $n$-node graph $\mathcal{G} =(\mathcal{V},\mathcal{E},\bm {W})$ (assuming it is an undirected and unweighted graph) is defined as 
    $\mathcal{V}^{\text{2hop}} = \mathcal{V}$,
    $(v_i,v_j) \in \mathcal{E^{\text{2hop}}}$ if the shortest path distance between $v_i,v_j$ in graph $\mathcal{G}$ is $2$ for $i,j\in[n]$, and 
\begin{gather*}
    \bm{W}^{\text{2hop}}_{i,j} =
    \begin{cases}
        1 & \text{if $(v_i,v_j) \in \mathcal{E_{\text{2hop}}}$,}\\
        0 & \text{otherwise,}
    \end{cases}
\end{gather*} 
for $i,j\in[n]$.
Based on the induced two-hop graph, we have the following definition, which is almost identical to Definition \ref{def_3}.

\begin{definition}\label{thf_def_3}
Let $\mathscr{T}_J(\mathcal G_{\mathrm{2hop}})$ be a multi-graph partition tree associated with an induced $2$-hop graph $\mathcal G_{\mathrm{2hop}}$ from $\mathcal{G}$. Let \{$\bm{A}^{[j,k]}$, $\bm{B}^{[j,k]}\setsep j=0,\ldots,J-1; k\in[n_j]\}$ be the filter matrices satisfying the assumptions of Theorem \ref{thm_1}. Then for  $J_0=0,\ldots, J-1$, the $\mathcal V^{\mathrm{2hop}}$-framelet system $\mathcal F_{J_0}^J(\mathcal V^{\mathrm{2hop}})$ as in \eqref{def:vfmt} is a tight frame for $\mathbb{R}^n$, which we call  a \textbf{two-hop framelet (THF) system}. In such a case, we simply denote  $\mathcal F_{J_0}^J(\mathcal G_{\mathrm{2hop}}):=\mathcal F_{J_0}^J(\mathcal{V}^{\mathrm{2hop}})$. In particular, we define  for the special case $J_0=0$, the THF system $\mathcal F_0^J(\mathcal G_{\mathrm{2hop}})$ as 
\begin{equation}\label{def:THF}
\mathsf{THF}_J(\mathcal{G_{\mathrm{2hop}}}):=\mathcal F_0^{J}(\mathcal G_{\mathrm{2hop}})=\{\bm\Phi_0,\bm\Psi_0,\cdots,\bm \Psi_{J-1}\}.
\end{equation} 
\end{definition}

For simplicity, let $\mathbb{V}_0$ be one-dimensional and its basis vector be the constant unit-norm vector. To do so, we let each $\bm{A}^{[j,k]}$ be the constant unit-norm vector of its relevant size. As each $\bm{A}^{[j,k]}$ is decided, according to Lemma \ref{lm_1}, each $\bm{B}^{[j,k]}$ should be a tight frame of the vector space orthogonal to $\Span(\bm{A}^{[j,k]})$. Let $\tilde{\bm{B}}^{[j,k]}$ be the remaining $|\mathcal{C}_{j,k}|-1$ orthonormal vectors in the complement of $\Span(\bm{A}^{[j,k]})$, which can be computed using singular value decomposition. We construct a tight frame $\bm{F}^{[j,k]}$ in $\mathbb{R}^{|\mathcal{C}_{j,k}|-1}$ as the coefficient matrix for remaining orthonormal vectors. Then the $\bm{B}^{[j,k]} := \bm{F}^{[j,k]}\tilde{\bm{B}}^{[j,k]}$ will be our final $\bm{B}$ filters. In details, $\bm{F}^{[j,k]}$ is obtained via frame completion \cite{ji2022construction} given $3$ vectors of size $|\mathcal{C}_{j,k}|-1$ sampled from Gaussian distribution. In this way, the size of  $\bm{F}^{[j,k]}$ is $3(|\mathcal{C}_{j,k}|-1)\times (|\mathcal{C}_{j,k}|-1)$. Frame completion \cite{ji2022construction} provides the advantage of adjusting the number of framelets without having zero vectors.

\par Intuitively, each framelet in the two-hop framelet system encodes certain structural information. Moreover, the redundancy in frames results in various framelets being chosen. However, using all framelets will be excessive. Thus, it is necessary to filter the framelets and choose only a certain portion. To do so, we adopt a heuristic approach, in which we measure the variance of each framelet with respect to the normalized Laplacian of the original graph (see (\ref{ch4_eq_1})), and then sort the variance to produce a ``spectral display'' of the framelets. The general idea is to integrate two-hop structural information from the two-hop framelets and one-hop structural information from the original graph. Given the spectral display, we can identify the parts of the framelets that provide the most varying patterns. These varying patterns are the features that we expect to be informative and beneficial for node classification. The whole procedure will be presented and discussed in detail in the experiments.

\subsection{Experiments}
To measure the extent of heterophily, we use the following quantity
\[
    H = \frac{\left|\{(u,v) \in \mathcal{E}|, y_u = y_v\}\right|}{|\mathcal{E}|}
\]
where $y_u,y_v$ are the labels of $u,v$. A smaller $H$ indicates a smaller portion of edges that connect nodes from the same classes, and thus a larger heterophily. We conducted experiments on three graphs, \textit{amazon-ratings}, \textit{minesweeper}, and \textit{tolokers}, from \cite{platonovcritical}. According to the measure above, \textit{amazon-ratings} possesses relatively larger heterophily, while the other two graphs have small heterophily and could be considered as homophilous graphs. However, as argued in \cite{platonovcritical}, the measure above has certain drawbacks and has high heterophily using other measures. Our proposal of including the graphs \textit{minesweeper} and \textit{tolokers} is to further demonstrate the effectiveness of our method in deriving rich structural information, not necessarily restricted to heterophilous graphs.

\par
\begin{figure*}[htp]
\centering

\subfloat[]{\includegraphics[width=0.25\textwidth]{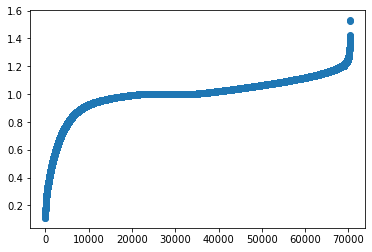}}
\subfloat[]{\includegraphics[width=0.25\textwidth]{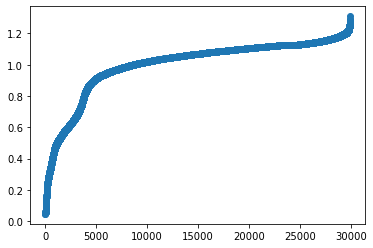}}
\subfloat[]{\includegraphics[width=0.25\textwidth]{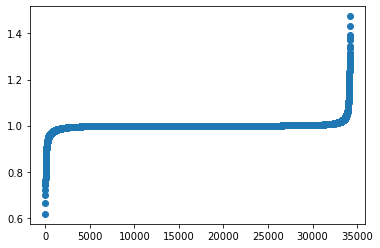}}
\caption{Distributions of framelet variance. (a): 70299 framelets on \textit{amazon-ratings}; (b): 29899 framelets on \textit{minesweeper};  (c): 34185 framelets on \textit{minesweeper}. Sorted non-decreasingly according to $\bm{f}\tilde{\bm{L}}\bm{f}^\top$.}
\label{thf_fig_1}
\end{figure*}

  In forming the partition tree, we controlled that each $|\mathcal{C}_{j,k}|\leq 16$. Then the $\mathsf{THF}_J(\mathcal{G}_{\mathrm{2hop}})$ was formed as described in Def. \ref{thf_def_3}. Each framelet was further normalized to have unit norm and sorted non-increasingly according to $\bm{f}\tilde{\bm{L}}\bm{f}^{\top}$. The figures in Figure \ref{thf_fig_1} show the overall distribution of variance of the framelets. These are the spectral displays we mentioned in the previous section. We can see that the first and final parts of the diagrams show drastic changes in variance. We considered these parts to have rich patterns and structural information due to the changes in variance. As a result, these framelets were selected as additional features to be input into neural networks. The implementations\footnote{https://github.com/zrgcityu/Framelets-as-features} were adapted to handle large graphs using sparse matrices.

\par
\begin{figure*}[htp]
\centering
\subfloat[top l]{\includegraphics[width=0.25\textwidth]{figures/az.png}}
\subfloat[top r]{\includegraphics[width=0.25\textwidth]{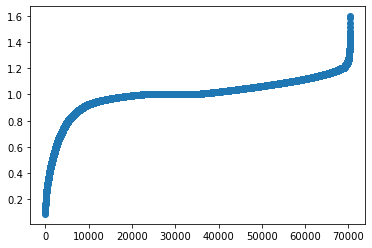}}
\subfloat[bottom l]{\includegraphics[width=0.25\textwidth]{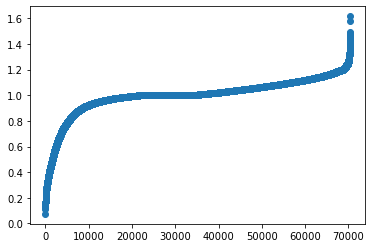}}
\subfloat[bottom r]{\includegraphics[width=0.25\textwidth]{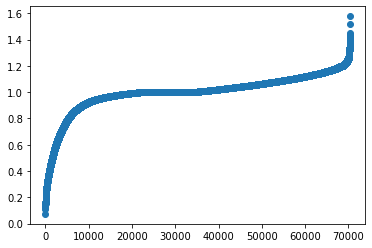}}
\caption{Distributions of framelet variance under different random seeds. Accuracy on $\textit{amazon-ratings}$: (a) 54.48 $\pm$ 0.27; (b): 54.40$\pm$0.41; (c): 54.31$\pm$0.51; (d): 54.10 $\pm$ 0.43.}
\label{thf_fig_2}
\end{figure*}

 To combine with GNNs, we selected certain framelets with either the lowest or the highest variance in the spectral displays and concatenated them with the original node feature matrices. Our results were obtained based on the best GNNs reported in \cite{platonovcritical}. In detail, for \textit{amazon-ratings}, based on a $2$-layer SAGE, we tested framelets with the lowest variance, and the number of framelets takes values in the set $\{1800,1900,2000,2100,2200\}$; for \textit{minsweeper} and \textit{tolokers}, based on a $5$-layer GAT-sep, we tested framelets with the highest variance, and the number of framelets takes values in the set $\{100,200,300,400,500\}$. The final results were chosen according to the performance on the validation sets. We adopted the publicly released code from \cite{platonovcritical} for the GNNs experiments. The results are shown in Table \ref{thf_t_1}, in which the results other than ours are cited from \cite{platonovcritical}. We can see that for all datasets, combined with our selected framelets as additional features, the GNNs were able to gain a certain amount of improvement and achieve the best results compared with all other methods. We further tested the stability in terms of randomness in generating framelets. As mentioned in Section \ref{gnn_subsec_1}, the generation of framelets involves random sampling. Figure \ref{thf_fig_2} shows that for four different random seeds, the framelets resulted in very similar performance, all of which outperformed the second-best result in Table \ref{thf_t_1}.

\begin{table}[htpb!]
	\scriptsize
	\centering
	\caption{Dataset statistics and classification results. Metric: \textit{amazon-ratings}, accuracy; \textit{minesweeper}, \textit{tolokers}, ROC AUC. \textbf{Best} with bold, \underline{second-best} with underline.}

	\label{tab:D}
	\begin{tabular}{@{}rccc@{}}
		\toprule
		& \textit{amazon-ratings} & \textit{minesweeper} & \textit{tolokers}\\ \midrule
Node                  &24492 &10000 &  11758\\
Feature              &300 &7  & 10 \\
Edge 	                & 93050 & 39402 & 519000 \\	
Class                &5 & 2 &2    \\
$H$     & 0.38 &  0.68 & 0.59\\ \midrule
ResNet \cite{he2016deep} & 45.90 $\pm$ 0.52 & 50.89 $\pm$ 1.39 & 72.95 $\pm$ 1.06\\
ResNet+SGC & 50.66 $\pm$ 0.48 & 70.88 $\pm$ 0.90 & 80.70 $\pm$ 0.97\\
ResNet+adj & 51.83 $\pm$ 0.57 & 50.42 $\pm$ 0.83 & 78.78 $\pm$ 1.11\\ \midrule
H2GCN \cite{zhu2020beyond}  & 36.47 $\pm$ 0.23 & 89.71 $\pm$ 0.31 &73.35 $\pm$ 1.01 \\
CPGNN \cite{zhu2021graph} & 39.79 $\pm$ 0.77 & 52.03 $\pm$ 5.46 & 73.36 $\pm$ 1.01 \\
GPR-GNN \cite{chien2021adaptive} & 44.88 $\pm$ 0.34 & 86.24 $\pm$ 0.61 & 72.94 $\pm$ 0.97 \\
FSGNN \cite{maurya2022simplifying} & 52.74 $\pm$ 0.83 & 90.08 $\pm$ 0.70 & 82.76 $\pm$ 0.61 \\
GloGNN \cite{li2022finding} &  36.89 $\pm$ 0.14 & 51.08 $\pm$ 1.23 & 73.39 $\pm$ 1.17 \\
FAGCN \cite{li2022finding} & 44.12 $\pm$ 0.30 & 88.17 $\pm$ 0.73 & 77.75 $\pm$ 1.05 \\
GBK-GNN \cite{du2022gbk} & 45.98 $\pm$ 0.71 & 90.85 $\pm$ 0.58 & 81.01 $\pm$ 0.67 \\
JacobiConv \cite{wang2022powerful} & 43.55 $\pm$ 0.48 & 89.66 $\pm$ 0.40 & 68.66 $\pm$ 0.65 \\ \midrule 
GCN \cite{kipf2016semi} &48.70 $\pm$ 0.63 & 89.75 $\pm$ 0.52 & 83.64 $\pm$ 0.67 \\
SAGE \cite{hamilton2017inductive} & \underline{53.63 $\pm$ 0.39} & 93.51 $\pm$ 0.57 & 82.43 $\pm$ 0.44 \\
GAT \cite{velivckovic2018graph} & 49.09 $\pm$ 0.63 & 92.01 $\pm$ 0.68 & 83.70 $\pm$ 0.47\\
GAT-sep & 52.70 $\pm$ 0.62 & \underline{93.91 $\pm$ 0.35} & \underline{83.78 $\pm$ 0.43} \\
GT \cite{shi2020masked} & 51.17 $\pm$ 0.66 & 91.85 $\pm$ 0.76 & 83.23 $\pm$ 0.64 \\
GT-sep & 52.18 $\pm$ 0.80 & 92.29 $\pm$ 0.47 & 82.52 $\pm$ 0.92 \\ \midrule
Ours & \textbf{54.48 $\pm$ 0.27} & \textbf{94.24 $\pm$ 0.44} & \textbf{84.38 $\pm$ 0.57}
\\
\bottomrule
	\end{tabular}
	\label{thf_t_1}
\end{table}

\section{Conclusion and Discussion}\label{sec:conclude}
We proposed a general system to generate frames on graphs that provides a practical solution to the following problem: sparsely representing arbitrary families of graph signals that are approximately low-dimensional. Our system is particularly suitable for the problem, as we proposed to learn the filters that adapt to the given signals. Apart from graph signal processing, we have shown that the system is also capable of generating rich and novel structural information that is beneficial to node classifications. However, the approach of selecting framelets for node classification is overall heuristic. In practice, this requires ad hoc and possibly a large number of attempts in choosing the framelets. It would be interesting to investigate whether there is any theoretical discussion or even guarantees for our framelet systems, as well as the ways of choosing framelets. 

The training time consumed in the Riemannian optimizations could be an issue. We would like to point out that some of the filters $\bm{A}^{[j,k]}$ can be predefined to reduce the number of parameters and, consequently, the training overhead. In this case, the system has a smaller representation capacity. Specifically, a detailed investigation of the computational scalability of the proposed framelet systems would be necessary for big-data scenarios. As it is proposed to conduct certain optimizations on Stiefel manifolds in a specific manner, an analysis concerning the choice of algorithms is beneficial, and it is also interesting to consider exploiting the iteration patterns and backpropagation in GPU computing, see e.g. \cite{xiao2025cdopt}.

As for further applications, given a pre-trained system, the fast forward and backward transforms resemble the computation in general neural networks and can be combined with a properly defined neural network. Using its sparseness, a pre-trained system can serve as a regularizer in deep learning tasks. It can also be applied in compressed sensing. These remain future directions to be explored.

Finally, our method can be applied to more general discrete and irregular data as long as a series of partitions is given. Thus, apart from graph data, it has the potential to be applied in other settings.




\end{document}